\documentclass[lettersize,journal]{IEEEtran}
\usepackage{amsmath,amsfonts}
\usepackage{algorithmic}
\usepackage{algorithm}
\usepackage{array}
\usepackage{textcomp}
\usepackage{stfloats}
\usepackage{url}
\usepackage{verbatim}
\usepackage{graphicx}
\usepackage{cite}
\usepackage{}
\usepackage{pdfpages}
\usepackage{tabularx}
\usepackage{multirow}
\usepackage{pgfplots}
\usepackage{subfigure}
\begin{document}

\title{Low Rank Properties for Estimating Microphones Start Time and Sources Emission Time}

\author{Faxian Cao, Yongqiang Cheng, Adil Mehmood Khan, Zhijing Yang, S. M. Ahsan Kazmi,  and Yingxiu Chang
        % <-this % stops a space
\thanks{This work was supported by China Scholarship Council ({\em Corresponding author: Yongqiang Cheng}).}% <-this % stops a space
\thanks{F. Cao, Y. Cheng, A. M. Khan and Y. Chang are with School of Computer Science, University of Hull, Hull HU6 7RX, U.K. (e-mail: faxian.cao-2022@hull.ac.uk; y.cheng@hull.ac.uk; a.m.khan@hull.ac.uk; y.chang-2020@hull.ac.uk).}

\thanks{Z. Yang is with School of Information Engineering, Guangdong University of Technology, Guangzhou 510006, China (e-mail:yzhj@gdut.edu.cn).}

\thanks{S. M. Ahsan Kazmi is with the Faculty of Computer Science and Creative
Technologies, University of the West of England, Bristol BS16 1QY, U.K.
(e-mail: ahsan.kazmi@uwe.ac.uk).}
}

\markboth{ Submitted to IEEE Transactions on xxx
%IEEE TRANSACTIONS ON SIGNAL PROCESSING
}%
{Shell \MakeLowercase{\textit{et al.}}: A Sample Article Using IEEEtran.cls for IEEE Journals}


\maketitle

\begin{abstract}
Uncertainty in timing information pertaining to the start time of microphone recordings and sources' emission time pose significant challenges in various applications, such as joint microphones and sources localization. Traditional optimization methods, which directly estimate this unknown timing information (UTIm), often fall short compared to approaches exploiting the low-rank property (LRP). LRP encompasses an additional low-rank structure, facilitating a linear constraint on UTIm to help formulate related low-rank structure information. This method allows us to attain globally optimal solutions for UTIm, given proper initialization. However, the initialization process often involves randomness, leading to suboptimal, local minimum values.
This paper presents a novel, combined low-rank approximation (CLRA) method designed to mitigate the effects of this random initialization. We introduce three new LRP variants, underpinned by mathematical proof, which allow the UTIm to draw on a richer pool of low-rank structural information. Utilizing this augmented low-rank structural information from both LRP and the proposed variants, we formulate four linear constraints on the UTIm. Employing the proposed CLRA algorithm, we derive global optimal solutions for the UTIm via these four linear constraints.
Experimental results highlight the superior performance of our method over existing state-of-the-art approaches, measured in terms of both the recovery number  and reduced estimation errors of UTIm.
\end{abstract}


\section{Introduction}

\IEEEPARstart{T}{he} accurate estimation of time of arrival (TOA) and time difference of arrival (TDOA)~\cite{1,2} measurements is a critical prerequisite for various applications, such as microphone array calibration~\cite{new1}, source localization and tracking~\cite{new6, new8, new9}, joint microphones and sources localization~\cite{29, 30, 31}, and simultaneous localization and mapping~\cite{new10, new11}. Specifically, the TOA measurements between individual microphones and sources are obtained through cross-correlation when the waveform of the source signal is known beforehand, such as its amplitude, frequency, and duration~\cite{4}. Additionally, the TDOA of a pair of microphones with respect to the corresponding source are determined by performing cross-correlation on the two corresponding microphone signals when the waveform of the source signal is unknown~\cite{gccp}. When microphones and sources are synchronous, meaning their recording start time and emission time align, the time difference between when an audio signal is emitted by the source and received by a microphone is measured using TOA~\cite{new4}. Similarly, TDOA measures the delay between a pair of microphones receiving corresponding audio signals~\cite{33}. However, these estimations are often complicated by unknown timing information (UTIm) stemming from asynchronous microphones and sources. Thus, the lack of synchronization in the recording start time of microphones and the emission time of sources significantly challenges the accurate estimation of timing information~\cite{8}.

Existing methods~\cite{ 6, 7, 8} to estimate UTIm in TOA/TDOA measurements primarily fall into two categories. The first category relies on direct optimization methods, including maximum likelihood estimate~\cite{6} and distributed damped Newton optimization~\cite{7}. Despite their popularity, these methods require both TDOA and angle of arrival information~\cite{22} and can be limited in certain scenarios. The second category~\cite{8} hinges on exploiting the low-rank property (LRP)~\cite{9}, where the low-rank structure information of UTIm is used to formulate linear constraints for UTIm estimation. The effectiveness of this method is often hindered by unknown and random initialization of UTIm, leading to local (suboptimal) minimal values and thus limited recovery rate (the ratio of number for successful initializations to all initializations in one configuration) and convergence rate (the ratio of number for successful recovery configurations to number of all configurations), particularly when there is noise in TOA measurements.

Drawing upon our prior work~\cite{3-1}, which highlighted the sufficiency of microphone signals alone for localizing microphones and sources in asynchronous environments, we delve deeper into the relationship between TOA and TDOA formulas in our previous work~\cite{3-1} by presenting a novel mapping function. More specifically, expanding on the significant alignment between the transformations of TOA and TDOA measurements discovered in our previous research, we imply that the low-rank structure information exploited between UTIm and TOA could be similarly applied to TDOA, suggesting the distances between microphones and sources can be acquired once the  UTIm in TOA or TDOA is estimated.

%presence of unexplored low-rank structure information between UTIm and TDOA that could enhance our UTIm estimation methodologies.

To this end, we introduce the combined low-rank approximation (CLRA) method to estimate UTIm in TOA/TDOA. This method integrates the linear constraints formed by LRP with three new variants of LRP designed to exploit more low-rank structure information between UTIm and TOA/TDOA. Then, by using these four linear constraints to restrict UTIm, our CLRA method seeks a global optimization solution for UTIm estimation, improving both convergence and recovery rates while reducing estimation errors in noise environment. In addition, alongside presenting this novel method, we provide a mathematical proof for the proposed LRP variants, reinforcing the theoretical foundation of our approach. This paper, therefore, not only uncovers a novel methodology for UTIm estimation but also presents robust evidence to validate its effectiveness. This study thus serves as a significant advancement in the realm of signal processing and localization, paving the way for more accurate and efficient methods for estimating UTIm in TOA/TDOA.

\section{State of The Art}

\begin{table*}[!t]
\begin{center}

\caption{Summary of State of The Arts for Estimating Unknown Timing Information in TOA or TDOA (ST: start time; ET: emission time; TOA: time of arrival; TDOA: time difference of arrival; AOA: angle of arrival; LRP: low-rank property; LRPV1, LRPV2 and LRPV3: proposed three variants of LRP; CLRA: combined low-rank approximation).}
\label{tablesum}
\begin{tabular}{p{1.2cm}|p{2.4cm}|p{0.5cm}|p{4.2cm}|p{3.3cm}| p{1cm}}
\hline
  \multirow{2}{*}{Reference} &  \multirow{2}{*}{Information Used} &  \multicolumn{2}{c|}{\multirow{2}{*}{Methods}}& \multicolumn{2}{c}{Prior Information} 
 \\ 
 \cline{5-6} 
 &  &\multicolumn{2}{c|}{}  & Constraints & Parameters \\  
\hline
 %\cite{10,11} & mic signal coherence & \multicolumn{2}{c|}{noise coherence match} & diffuse noise field & \multirow{3}{*}{CT}  \\ \cline{1-5} 
 
%  \cite{24}& \multirow{2}{*}{TDOA}	&\multicolumn{2}{c|}{rank-5
%factorization} & \multirow{2}{*}{--}   &	 \\  \cline{1-1} \cline{3-4}  

 %\cite{31} &	&  \multicolumn{2}{c|}{\multirow{3}{*}{LRP; SVD} }  & 		&  \\ \cline{1-2}  \cline{5-6}  

 %  \cite{4} & \multirow{5}{*}{TOA}	&  \multicolumn{2}{c|}{}	 & co-located mic and source &  \multirow{3}{*}{CT; OT}		 \\  \cline{1-1} \cline{5-5}

 %\cite{29,30} &		&  \multicolumn{2}{c|}{}  	& \multirow{3}{*}{--} 	&   \\ \cline{1-1} \cline{3-4} 

 % \cite{26, 27, 28} &   &	\multicolumn{2}{c|}{Grobner basis}	&    & 	  \\\cline{1-1} \cline{3-4}  \cline{6-6}

 \cite{32}&  \multirow{2}{*}{TOA}	& \multicolumn{2}{c|}{probabilistic generative model} &	  --& ST	 \\   \cline{1-1} \cline{3-6}

  \cite{34}	& 	& \multicolumn{2}{c|}{Gram matrix}	& known center of geometry	& \multirow{6}{*}{--}  \\  \cline{1-5} 

% \cite{15,16,17,18} & TDOA & \multicolumn{2}{c|}{min/max TDOA estimation} & end-fire sources &   \\
%\cline{1-5}
%  \cite{19,20,21} & mic signal energy & \multicolumn{2}{c|}{energy-based method} & co-located mics and sources &  \\ 
%\cline{1-5}

 %\cite{22}& 	AOA &	\multicolumn{2}{c|}{rank property-AOA}	 & \multirow{3}{*}{near field assumption}  &	 \\ \cline{1-4}
 
   \cite{6} & \multirow{2}{*}{TDOA; AOA} &	\multicolumn{2}{c|}{maximum likelihood estimation} & \multirow{2}{*}{--} &	    \\ 
\cline{3-4} \cline{1-1}
 \cite{7} & 	 &	\multicolumn{2}{c|}{distributed damped Newton optimization} &	 &	 \\ \cline{1-5} 

 \cite{5} &	TOA/TDOA	& \multicolumn{2}{c|}{maximal likelihood estimation} &	co-located mics and sources & 	 \\\cline{1-2} \cline{3-5}

 \cite{33} & \multirow{2}{*}{TDOA} 		& \multicolumn{2}{c|}{auxiliary function approach}	& \multirow{6}{*}{--} 	&  \\ \cline{1-1} \cline{3-4} 

 \cite{37,38} & 	 & \multirow{5}{*}{LRP} 	& nuclear truncation minimization &	 &  \\  \cline{4-4}  \cline{1-2} \cline{6-6} 
 
  %\cite{3} &   & & \multirow{2}{*}{structure total least square}   &  &	  \\  \cline{1-2}  \cline{6-6} 
  
   \cite{8,39} & \multirow{2}{*}{TOA} 	& & structure total least square &	 	&  \multirow{2}{*}{ET}  \\ \cline{1-1}  \cline{4-4}

 \cite{36}	&		&  & \multirow{2}{*}{alternating minimization} 	& 	&  \\  \cline{1-2}
\cline{6-6}

 \cite{35}	& & &  	& 	&   \multirow{2}{*}{--} \\ \cline{1-2} \cline{4-4}

 Proposed	& TOA/TDOA	 & & LRPV1;
LRPV2; LRPV3; CLRA&	 	&    \\ \hline
\end{tabular}
\end{center}
\end{table*}

This section provides a review of the state-of-the-art methods used for estimating UTIm in TOA/TDOA measurements as shown in Table \ref{tablesum}. As stated early in Section I, the methods for UTIm estimation are categorized into two groups, i.e., direct optimization and LRP-based~\cite{9}. Among the direct optimization techniques, probabilistic generative model~\cite{32} estimates sources emission time by using TOA measurements. However, this model ignores the start time of microphones in TOA measurements. To eliminate UTIm in TOA measurements, one work introduces Gram matrices~\cite{34}, assuming that the center point among microphones and sources is known. Nonetheless, this assumption does not hold for microphones array applications. Maximum likelihood estimation~\cite{5} assumes co-located sources and microphones and estimates their locations and microphone start time using either TOA or TDOA information. In addition, optimal solutions are derived through close form equations in this case~\cite{5}. Auxiliary function-based approach~\cite{33} uses TDOA information to estimate the locations and microphone time offset, showing better convergence properties. Some approaches, such as maximum likelihood estimation~\cite{6} and distributed damped Newton optimization~\cite{7}, combine TDOA with AOA measurements to estimate UTIm. However, these approaches require both TDOA and AOA measurements and this is not usually satisfied in many scenarios. Among the first group of methods reviewed, only auxiliary function approach~\cite{33} presents a general method while others require additional geometric constraints on the location of microphones and sources or assume the start or emission time to be known or necessitate more measurements.

Regarding the LRP-based methods for estimating UTIm~\cite{9}, they are categorized into three algorithms: alternating minimization (AM)~\cite{35,36}, nuclear truncation minimization (NTM)~\cite{37,38}, and structure total least square (STLS)~\cite{8,39}. Of these algorithms, the STLS ~\cite{8,39} method is over 100 times faster than both AM and NTM. %Additionally, 
%one study ~\cite{36} used the AM method, while two studies used the STLS ~\cite{8,39}, however, the emission time of sources was set to 0.
Besides, by denoting the time offset of the microphones in TDOA as the pseudo-start time of the microphones and the distance (divided by the speed of sound) between the first microphone and the corresponding source as the pseudo-source emission time~\cite{34}, the LRP~\cite{9} is being used to estimate UTIm with TDOA information. In our previous work~\cite{3-1}, we demonstrated that the transformations of both TOA and TDOA measurements are identical to one another, showing the UTIm in TOA and TDOA are the same as each other. Therefore, the values of variable in the low-rank matrix exploited by LRP with TDOA measurements can be equivalent to those in the low-rank matrix exploited by LRP with TOA information.

In summary, various methods are proposed to estimate UTIm using TOA or TDOA measurements. Among these methods, the auxiliary function-based algorithm~\cite{33} and LRP-based methods~\cite{35} are promising due to their consideration for a general formulation. However, LRP-based~\cite{9} methods are found to have better recovery and convergency rates with less estimation errors of UTIm than the auxiliary function-based algorithm~\cite{33}. In addition, by using LRP~\cite{9}, STLS~\cite{8,39} is found to be 100 times faster than AM~\cite{35, 36} and NTM~\cite{37,38}. Nevertheless, with LRP only, resulting in a risk of solutions getting stuck in local minimal values. To address this issue, this paper presents three variants of LRP that exploit more of the low-rank structure information between TOA/TDOA and UTIm. The proposed CLRA uses both LRP and the corresponding three new variants to constrain UTIm, allowing for a global solution to be found. Therefore, the proposed CLRA method helps improve the recovery and convergency rate  while also reducing estimation errors of UTIm.

\section{Problem Formulation}
There are $M$ microphones and $N$ sources located at unknown locations $R=\begin{bmatrix}
    r_1, & r_2, & \cdots, & r_M
\end{bmatrix}_{3 \times M}$ and $S= \begin{bmatrix}
    s_1, & s_2, & \cdots, & s_N
\end{bmatrix}_{3\times N}$, respectively, where $3$ represents the 3 dimensional space. Then, if there is a fundamental control centre for both microphones and sources, the microphones and sources can be synchronized, i.e., both the emission time of sources and the microphones start time are known. Thus TOA ($t_{i,j}=\frac{\|r_i-s_j\|}{c}$) between $i^{th}$ microphone and $j^{th}$ source is obtained with the received microphone signal and known waveform of source signal~\cite{4} (Note that $\|\bullet\|$ and $c$ are the $l_2$ norm and speed of sound, respectively.) Similarly, TDOA ($\zeta_{i,j}=\frac{\|r_i-s_j\|-\|r_1-s_j\|}{c}$) for a pair of microphones with respect to $j^{th}$ source is obtained using the received signals from a pair of microphones~\cite{gccp}. However, usually the microphones and sources are asynchronous~\cite{3-1, 34, 39}, resulting in the emission time of sources and start time of microphones being unknown.

%If the TOA $t_{i,j} = {\frac{\|r_i-s_j\|}{c}}$ between each pair of microphone and source $(r_i, s_j)$ is known (\(c\) is the speed of sound), the location of the microphone and sources can be easily estimated, for example, based on the existing rank property presented by \textit{Schönemann}~\cite{9}, \textit{Crocco et al.} presented bi-linear method to recover the location of microphones and sources by applying the SVD to the matrix which was constructed by the TOA.  However, this situation needs both the microphones and sources to be synchronized, 

If microphones and sources are asynchronous and the waveform of source signal is known, TOA between microphone and source is incomplete because of the UTIm. Denote $\delta=\begin{bmatrix}
   \delta_1, & \delta_2, & \cdots, &\delta_M
\end{bmatrix}^T$
 and $\eta=\begin{bmatrix}
   \eta_1, & \eta_2, & \cdots, &\eta_N
\end{bmatrix}^T$ where \(\delta_i\) and \(\eta_j\) are the unknown start time of $i^{th}$ microphone and unknown emission time of $j^{th}$ source, respectively (\(i\) and \(j\) range from 1 to \(M\) and 1 to \(N\), respectively). Then the TOA between microphones and sources is~\cite{34} \vspace{-0.2em}
\begin{equation}
\label{eq1}
t_{i,j}= {\frac{\|r_i-s_j\|}{c}}+\eta_j-\delta_i.
\vspace{-0.2em}
\end{equation}

If microphones and sources are asynchronous and the waveform of source signal is unknown, then TDOA is available and the received signals from a pair of microphones are utilized for TDOA estimation with the method of cross-correlation~\cite{gccp}.
Thus the relationship between TDOA and the location of microphones and sources is displayed as~\cite{34}
%\begin{align}
%\label{eq2}
 %   & \zeta_{i,j}=t_{i,j}-t_{1,j}= \frac{\|r_i-s_j\|-\|r_1-s_j\|}{c}+\delta_1-\delta_i \nonumber \\
 %   &  =\frac{\|r_i-s_j\|}{c}-\frac{\|r_1-s_j\|}{c}+\delta_i' ,
%\end{align}
\begin{equation*} 
\zeta_{i,j}=t_{i,j}-t_{1,j}= \frac{\|r_i-s_j\|-\|r_1-s_j\|}{c}+\delta_1-\delta_i
\vspace{-0.12cm}
\end{equation*}
\begin{equation}
\label{eq2}
   =\frac{\|r_i-s_j\|}{c}-\frac{\|r_1-s_j\|}{c}+\delta_i' ,
   \vspace{-0.12cm}
\end{equation}
where \(\delta_{i}^{'}\) is the time offset between  the $i^{th}$ microphone and  $1^{st}$ microphone.% and Fig. \ref{SignalModel} illustrates TOA formulation in Eq. (\ref{eq1}) and TDOA formulation in Eq. (\ref{eq2}).

Denote $\hat{\eta}_j=-\frac{\|r_1-s_j\|}{c}$ and $\hat{\delta}_i=-\delta_i'$, TDOA formula in Eq. (\ref{eq2}) is rewritten as~\cite{34}
\begin{equation}
    \label{andreaa_44}
    \zeta_{i,j}=\frac{\|r_i-s_j\|}{c}+\hat{\eta}_j-\hat{\delta}_i,
\end{equation}
which shares the same structure with the TOA formula in Eq. (\ref{eq1}).

By using TOA information $t_{i,j}$ only in Eq. (\ref{eq1}), the transformation of TOA formula in Eq. (\ref{eq1}) can be rewritten as~\cite{3-1}
\begin{equation}
    \label{andreaa_1}
      \dot{t}_{i,j} =\frac{\|r_i-s_j\|}{c}-\dot{\delta}_i+\dot{\eta}_j,  
\end{equation}
where $\dot{t}_{i,j}$ is the transformation of TOA $t_{i,j}$, and $\dot{\delta}_i$ and $\dot{\eta}_j$ are the transformation of microphone start time $\delta_i$ and source emission time $\eta_j$, respectively.

%\begin{figure}[t]
%\centering
%    \includegraphics[width=0.4\textwidth ]{SignalModel.jpg}
%\caption{The illustration~\cite{3} for the time of arrival formulation in Eq. (\ref{eq1}) and time difference of arrival formulation in Eq. (\ref{eq2}) with $j^{th}$ source and a pair of mics (%rectangular shape for source and mics denotes the emitted time of audio signal and received time of audio signal, i.e., the dash line for three rectangular shapes; 
%$r_1$: location of  $1^{st}$ mic; $r_i$: location of $i^{th}$ mic; $s_j$: location of $j^{th}$ source; $\eta_j$: emission time of $j^{th}$ source; $\delta_1$: start time of $1^{st}$ mic; $\delta_i$: start time of $i^{th}$ mic; $\delta_i'$: time offset between $1^{st}$ mic and $i^{th}$ mic; $t_{i,j}$: time of arrival between $i^{th}$ microphone and $j^{th}$ source; $t_{1,j}$: time of arrival between $1^{st}$ microphone and $j^{th}$ source; $\zeta_{i,j}$: time difference of arrival between $1^{st}$ microphone and $i^{th}$ microphone with respect to $j^{th}$ source; $c$: speed of sound; $\|\bullet\|$: 2 norm).}
%\label{SignalModel}
% \vspace{-1.6em}
%\end{figure}

Similarity,  by using TDOA information $\zeta_{i,j}$ only in Eq. (\ref{eq2}), the transformation of TDOA formula in Eq. (\ref{eq2}) is rewritten as~\cite{3-1}
\begin{equation}
    \label{andreaa_2}
     \ddot{\zeta}_{i,j} =\frac{\|r_i-s_j\|}{c}-\ddot{\delta}_i+\ddot{\eta}_j,
\end{equation}
where  $\ddot{\zeta}_{i,j}$ is the transformation of TDOA $\zeta_{i,j}$, and $\ddot{\delta}_i$ and $\ddot{\eta}_j$ are pseudo start time and pseudo emission time, respectively, and  \vspace{-0.15cm} 
\begin{equation}
     \label{andreaa_3}
     \begin{cases}
    \ddot{\eta}_j= \dot{\eta}_j \\
    \ddot{\delta}_i=\dot{\delta}_i \\
    \ddot{\zeta}_{i,j}=\dot{t}_{i,j} 
     \end{cases},
      \vspace{-0.15cm}
\end{equation}
where $i=\begin{matrix}
    1, & \cdots, & M
\end{matrix}$, $j=\begin{matrix}
    1, & \cdots, & N
\end{matrix}$ and $\ddot{\eta}_j= \dot{\eta}_j=0$.

Upon inspection of Eqs. (\ref{andreaa_1}), (\ref{andreaa_2}) and (\ref{andreaa_3}), it is obvious that the transformation of TDOA formula in Eq. (\ref{eq2}) can be same as the transformation of TOA formula in Eq. (\ref{eq1})~\cite{3-1}.
Without further mention, we denote $\dot{t}_{i,j}$ as TOA/TDOA, and also denote $\dot{\delta}_i$ and $\dot{\eta}_j$ as microphone start time and source emission time of TOA/TDOA, respectively.
%In practice, the TOA between emitter and microphone is obtained by the cross correlation method~\cite{3,39} with the corresponding emitted signal and received signal. If the source signal is unknown, the TDOA of a pair of microphones can be estimated by cross-correlation of the corresponding two microphone signals~\cite{3,39}. 
 Our objective is to estimate the UTIm from TOA/TDOA information.
 %And we will also prove the transformation of (\ref{eq2}) can be same as the transformation of (\ref{eq1}). It will lead the conclusion that source signal from sources for localization are not important any more.
 Due to the invariance of rotation and translation as well as reflection for the geometry of microphones and sources, i.e., the relative positions among microphones and sources, without loss of generality,  the locations for the first microphone and first source can be defined as $r_1=\begin{bmatrix}
     r_{1,1}, & 0, & 0
 \end{bmatrix}^T$ and $s_1=\begin{bmatrix}
    0, & 0, &0
 \end{bmatrix}^T$, respectively, where $r_{1,1}>0$. And the emission time for the first source can be set to be zero, i.e. \( \eta_{1}=0\)~\cite{35}. 

\section{Preliminaries}
 The low-rank structure information between UTIm and TOA/TDOA exploited by LRP~\cite{9} is stated at this section.
 
  By taking the square of both sides of Eq. (\ref{andreaa_1})~\cite{35}, we can have
\begin{align}
\label{eq3}
    & \frac{r_{i}^{T} r_{i}+s_{j}^{T} s_{j}-2r_{i}^{T} s_{j}}{c^{2}} \nonumber \\
   & =\dot{t}_{i,j}^2+\dot{\eta}_{j}^2 +\dot{\delta}_{i}^2
-2(\dot{t}_{i,j} \dot{\eta}_{j}-\dot{t}_{i,j} \dot{\delta}_{i} +\dot{\eta}_{j} \dot{\delta}_{i}),
\end{align}
where $i=\begin{matrix}
1, & \cdots, & M
\end{matrix}$ and  $j=\begin{matrix}
1, & \cdots, & N
\end{matrix}$.  

To formulate LRP, we can sequentially subtract the corresponding equation for $i=1$ and the equation for $j=1$ and then add the equation for $i=j=1$ from the general form Eq. (\ref{eq3}),  finally based on the assumption \(\eta_1=0\), it follows~\cite{35}
%
%we define other three formulations based on (\ref{eq3}): 1) let \textit{i}  be 1 while \textit{j} remains unchanged; 2) let the \textit{j} be 1 while  \textit{i} remains unchanged; 3) let both \textit{i} and \textit{j} be 1; Then (\ref{eq3}) subtracts first two equations and add the third equation in sequences, 
%
%
%
%By using the assumption \(\eta_1=0\) and from (\ref{eq3})
%
\begin{align}
	\label{eq4}
	    &  \frac{{-2(r_i-r_1)}^{T}(s_j-s_1)}{c^{2}} \nonumber \\  
	   & =\dot{t}_{i,j}^2-\dot{t}_{i,1}^{2}-\dot{t}_{1,j}^{2}+\dot{t}_{1,1}^{2}+2\dot{\delta}_i(\dot{t}_{i,j}-\dot{t}_{i,1}) \nonumber \\    
   &  -2\dot{\delta}_1(\dot{t}_{1,j}-\dot{t}_{1,1})-2\dot{\eta}_j (\dot{t}_{i,j}-\dot{t}_{1,j}) +2\dot{\eta}_j (\dot{\delta}_1-\dot{\delta}_i), 
\end{align}
%\begin{equation*}
    % \frac{{-2(r_i-r_1)}^{T}(s_j-s_1)}{c^{2}}=\dot{t}_{i,j}^2-t_{i,1}^{2}-t_{1,j}^{2}+t_{1,1}^{2}+2\delta_i(\dot{t}_{i,j}-t_{i,1})
%\end{equation*}
%\begin{equation}
%	\label{eq4}
 %   -2\delta_1(t_{1,j}-t_{1,1})-2\eta_j (\dot{t}_{i,j}-t_{1,j}) -2\eta_j (\delta_1-\delta_i), 
%\end{equation}
where \(i=2,\ \cdots,\ M\) and \(j=2,\ \cdots,\ N\). By defining four matrices \({R}\in \mathbb{R}^{3 \times (M-1)}\) , \({S}\in \mathbb{R}^{3 \times (N-1)}\), \(D\in \mathbb{R}^{(M-1) \times (N-1)}\) and \(U\in \mathbb{R}^{(M-1) \times (N-1)}\) as
\begin{align}
   R_{:,i-1}&=r_i-r_1, \nonumber \\
   S_{:,j-1}&=s_{j}-s_{1},  \nonumber \\
   D_{i-1,j-1}&=\dot{t}_{i,j}^2-\dot{t}_{i,1}^2-\dot{t}_{1,j}^2+\dot{t}_{1,1}^2, \nonumber\\
        U_{i-1,j-1}&=2\dot{\delta}_i\left(\dot{t}_{i,j}-\dot{t}_{i,1}\right)-2\dot{\delta}_1\left(\dot{t}_{1,j}-\dot{t}_{1,1}\right) \nonumber \\
       & -2\dot{\eta}_j\left(\dot{t}_{i,j}-\dot{t}_{1,j}\right)+2\dot{\eta}_j\left(\dot{\delta}_1-\dot{\delta}_i\right) \nonumber, 
  \end{align}
respectively,  where \(i=\begin{matrix}
  2,& \cdots,& M
  \end{matrix}\) and \(j=\begin{matrix}
  2,& \cdots,& N
  \end{matrix}\), then Eq. (\ref{eq4}) is written as matrix form~\cite{35}
\begin{equation}
   \label{eq5}
\frac{-2{R}^TS}{c^2}=D+U.
\end{equation}
%where \({\check{R}}\in \mathbb{R}^{3 \times (M-1)}\) , \({\check{S}}\in \mathbb{R}^{3 \times (N-1)}\), \(D\in \mathbb{R}^{(M-1) \times (N-1)}\) and \(U\in \mathbb{R}^{(M-1) \times (N-1)}\) can be represented as.
  
Upon inspection of Eq. (\ref{eq5}), the left side contains the information on the unknown location of microphones and sources while the right side contains  both TOA/TDOA and UTIm, it implies that we need to estimate the UTIm before localizing both microphones and sources. Then with Eq. (\ref{eq5}), the low-rank structure information between UTIm and TOA/TDOA exploited by LRP~\cite{9} is presented here.

{\bf LRP:}   if
\begin{equation}
\label{eeq7}
\begin{cases}
 M-1> 3 \\ N-1> 3
\end{cases},
\end{equation}
LRP can be stated as
\begin{equation}
\label{eq7}
 rank(D+U)= rank(R^TS)\leq 3.
\end{equation}
% \begin{equation}
 %\label{eq7}
% \le min{\left(rank\left(\check{R}\right),rank\left(\check{S}\right)\right)}=3.     
% \end{equation}

Upon inspection of Eq. (\ref{eq7}), matrix $U$ contains both the UTIm and TOA/TDOA, and matrix $D$ contains TOA/TDOA, it indicates that Eq. (\ref{eq7}) shows the low-rank structure information between the known TOA/TDOA and UTIm. However,
usually the initialization of UTIm is random, so that the solution of UTIm is easy to stuck into local minimal value. It leads the following problems: First, if the number of microphones or sources is not sufficient, the convergency rate   is limited, for example, if number of microphones is less than seven or number of sources is less than six, the convergency rate  achieved by LRP is almost zero percent. Second, the recovery rate  is still limited whatever the number of microphones and sources is. Third, the estimation error of UTIm tends to be large when there are noises in TOA/TDOA measurements. Therefore, we are interested in investigating whether there are any other low-rank structure information between UTIm and TOA/TDOA, i.e., given the additional linear constraints formulated by these low-rank structure information, we are aiming to improve  both the recovery and convergency rate of UTIm and reduce the estimation errors of UTIm  in noise environments.

\section{Proposed low-rank Properties}
This section presents proposed three variants of LRP that are used for estimating the UTIm. LRP shows the low-rank structure information between UTIm and TOA/TDOA, and the proposed three variants of LRP provide more low-rank structure information between UTIm and TOA/TDOA, thus with  all of these structure information, both the recovery and convergency rates for UTIm estimation is improved and estimation errors of UTIm is reduced in noise environments.

%Now, we will show one existing low-rank property from \cite{3, 9,39} first and we name this existing rank property as LRP. Then the proposed three new low-rank properties will be presented, and we name them as LRPV1, LRPV2 and LRPV3, respectively.
%In addition, it can be seen from \textit{Appendix A of Supplementary Material} that the transformation of (\ref{eq2}) with TDOA information can be the same as the transformation of (\ref{eq1}) with TOA information.
%\footnote{\redcolor{ps: I prepare to use a web-link to show the supplementary material, i.e. hang on them at arxiv}}

By defining three new matrices
\begin{equation}
  \label{thrm1}
T_1^\ast=
 \begin{bmatrix}
 D &  U
 \end{bmatrix} \in \mathbb{R}^{(M-1) \times 2(N-1)},
\end{equation}

\begin{equation}
  \label{thrm2}
 T_2^{\ast}=
 \begin{bmatrix}
 D^T &  U^T
 \end{bmatrix} \in \mathbb{R}^{(N-1) \times 2(M-1)},
\end{equation}
and
\begin{equation}
     \label{thrm3}
     T_3^\ast=
 \begin{bmatrix}
 D & U\\
 U & D
 \end{bmatrix}\in \mathbb{R}^{2(M-1) \times 2(N-1)}, 
\end{equation}
then three variants of LRP are proposed if Eq. (\ref{eeq7}) holds.

%Although this rank property has obtained good performance in terms of convergency rate when the number of microphone and source are large. The convergency rate is still quite low if \(M\) and \(N\) is small, in addition, the recovery rate is still quite low is quite high whatever \(M\) and \(N\) is large or small. 

% \subsection{LRPV1}
{\bf LRPV1:} %A variant of LRP can be proposed by defining a new matrix $T_1^*$ as
% \begin{equation}
 %\label{eq8}
 %T_1^\ast=
 %\begin{bmatrix}
 %D &  U
 %\end{bmatrix} \in \mathbb{R}^{(M-1) \times 2(N-1)},
 %\end{equation} 
%and
%If (\ref{eeq7}) holds, we have
% the number of rank for matrix \(T_1^\ast\) equals to \(N-1+3\) once (\ref{eeq7}) and \( M-1>N-1+3\) hold, in other words,
\begin{equation}
\label{eq9}
  rank\left(T_1^\ast\right)\leq min \{M-1, N-1+3\},
  %\begin{cases}
  % N-1+3 & \text{if } M-1>N-1+3, \\
 %  M-1 & \text{otherwise},
%  \end{cases}
\end{equation}
where matrix $T_1^\ast$ is low-rank only if $ M-1>N-1+3$
(see proof in \textit{Section A1 of Supplementary Material}).

% and this is the proposed rank property. 

%\textit{Theorem 1.1}  and \textit{1.2} reveals one new algebraic property of TOA and can be applied to TDE. And we can see that the \textit{Theorem 1.1} is same as \textit{Theorem 1.2} when \(N-1\) equals to \(3+1\). We, hence, can regard \textit{Theorem 1.2} as the general case for \textit{Theorem 1.1}. Compared  with the existing rank property proposed by \textit{Schönemann} which one precondition needs to be held (\(\big\{\begin{matrix}M> 3+1 \\ N> 3+1 \end{matrix}\)), we can see that the \textit{Theorem 1.2} has two preconditions for holding the low-rank property, they are \(\big\{\begin{matrix}M> 3+1 \\ N> 3+1 \end{matrix}\) and \( M>N+3\), respectively.

{\bf LRPV2:} 
%A variant of LRP can be proposed by defining a new matrix $T_2^*$ as
% \begin{equation}
% \label{eq10}
%T_2^{\ast}=
% \begin{bmatrix}
% D^T &  U^T
% \end{bmatrix} \in \mathbb{R}^{(N-1) \times 2(M-1)},
% \end{equation}
%and 
%If (\ref{eeq7}) holds, we have
%It has been proved in the \textit{Appendix C} that for any positive integer \(3+1\le M-1\), the rank of matrix \(T_2^\ast\)  equals to \(M-1+3\)  once (\ref{eeq7}) and \(N-1>M-1+3\) hold, in other words,                  
\begin{equation}
\label{eq11}
    rank\left(T_2^\ast\right)\leq min \{N-1, M-1+3\},
\end{equation}
where matrix $T_2^\ast$ is low-rank only if $ N-1>M-1+3$
(see proof in \textit{Section A2 of Supplementary Material}).

%The size of matrix $T_2^\ast$  is $(N-1) \times 2(M-1)$, thus, if  (\ref{eeq7}) and \( N-1>M-1+3\) hold, then (\ref{eq11}) also reveals a low-rank property and this is the proposed rank property. 
 
%The second rank property of TOA can be revealed by \textit{Theorem 2.1}  and \textit{2.2} and can be applied to TDE. And we can also see that the \textit{Theorem 2.1} is same as \textit{Theorem 2.2} when \(M-1\) equals to \(3+1\). We, hence, can regard \textit{Theorem 2.2} as the general case for \textit{Theorem 2.1}. Similar with the \textit{Theorem 1.2}, we can see that \textit{Theorem 2.2} has an additional precondition, that's \( N>M+3\). 

{\bf LRPV3:}
%A variant of LRP can be proposed by defining a new matrix $T_3^*$ as
%\begin{equation}
%\label{eq12}
%    T_3^\ast=
% \begin{bmatrix}
% D & U\\
% U & D
% \end{bmatrix}\in \mathbb{R}^{2(M-1) \times 2(N-1)},
%\end{equation}
%and 
%If (\ref{eeq7}) holds, we have
%It has been proved in the \textit{Appendix D} that for any positive integer \(3\) if (\ref{eeq7}) holds, the number of rank for matrix \(T_3^\ast\)  equals to $min (N-1+3, M-1+3)$, in other words,
\begin{equation}
\label{eq13}
  rank\left(T_3^\ast\right)\leq min \{N-1+3,M-1+3\},
\end{equation}
where matrix $T_3^\ast$ is always low-rank if Eq. (\ref{eeq7}) holds (see proof in \textit{Section A3 of Supplementary Material}). 

From Eqs. (\ref{thrm1}), (\ref{thrm2}), (\ref{thrm3}), (\ref{eq9}), (\ref{eq11}) and (\ref{eq13}), we can see that LRPV1, LRPV2 and LRPV3 always reveal low-rank structure information between known TOA/TDOA and UTIm if $M-1>N-1+3$ holds for $T_1^\ast$ and $N-1>M-1+3$ holds for $T_2^\ast$.

%that once $M-1>N-1+3$ holds for LRPV1 and $N-1>M-1+3$ holds for LRPV2, both LRPV1 and LRPV2 can reveal low-rank structure information between known TOA/TDOA and unknown timing information since both $T_1^\ast$ and $T_2^\ast$ contain them. In addition, From (\ref{eq13}), we can also see that LRPV3 always has low-rank structure information between TOA/TDOA and unknown timing information.    

%Once $M-1>N-1+3$, LRPV1 can also reveal low-rank structure information between known TOA/TDOA and unknown timing information since $T_1^\ast$ contains them.

%Once $N-1>M-1+3$, LRPV2 can also reveal low-rank structure information between known TOA/TDOA and unknown timing information since $T_2^\ast$ contains them.

%LRPV3 can also reveal low-rank structure information between known TOA/TDOA and unknown timing information since $T_3^\ast$ contains them. 

%In addition, Fig. \ref{figurefor4proeprtyies} illustrated the precondition that those four Properties above has low-rank structure information between unknown timing information and TOA/TDOA. \vspace{-0.1em}
%
%The size of matrix $T_3^\ast$  is $2(M-1) \times 2(N-1)$, thus, if   (\ref{eeq7}) holds, then (\ref{eq13}) also reveals a low-rank property and this is the proposed rank property. 

\section{Proposed Algorithm}
This section illustrates the corresponding linear constraints first based on low-rank structure information exploited by both LRP in Section IV and three variants of LRP in Section V, and then applies them to UTIm estimation. %In common situations, all microphones and sources lie in the 3D space, thus we set \(3=3\).

\subsection{Linear constraint based on LRP}
From Eqs. (\ref{eq5}) and (\ref{eq7}), denote 
\begin{equation}
    \label{eq14}
    \begin{cases}
     D=\begin{bmatrix}
         A & B
     \end{bmatrix} \\
     U=\begin{bmatrix}
         F & G
     \end{bmatrix}
    \end{cases},
\end{equation}
%\begin{align}
% \label{eq14}
 %   & D=[A \ B], \nonumber\\
 %    &U=[F \ G],
%\end{align}
 %  \begin{equation}
 %  \label{eq14}
 %  \begin{cases}
 %   D=\begin{bmatrix}
  %   A & B
  %  \end{bmatrix}  \\
  %     U=\begin{bmatrix}
 %    F & G 
  %  \end{bmatrix}
 %  \end{cases}
%\end{equation}
%
%\begin{equation}
%\label{eq14}
%    U=\begin{bmatrix}
%     F & G 
 %   \end{bmatrix}
%\end{equation}
where \(A\in \mathbb{R}^{(M-1) \times 3}\), \(B\in \mathbb{R}^{(M-1) \times (N-1-3)}\), \(F\in \mathbb{R}^{(M-1) \times 3}\) and \(G\in \mathbb{R}^{(M-1) \times (N-1-3)}\). Then we have
\begin{equation}
\label{eq16}
rank(\begin{bmatrix}
     A+F & B+G
\end{bmatrix})=rank(A+F) \leq 3.
\end{equation}

From Eq. (\ref{eq16}), we can assume there is a matrix $X$ that enables 
    \begin{equation}
    \label{eq15}
        (A+F)X=B+G,
    \end{equation}
where $X\in \mathbb{R}^{3 \times (N-1-3)}$ is unknown and is estimated in Section VI-E.
%where $X \in \mathbb{R}^{3 \times (N-1-3)}$
%\begin{equation*}
 %   X=\begin{bmatrix}
  %    X_{1,1} & \cdots &  X_{1,N-1-3}\\
  %   X_{2,1} & \cdots &  X_{2,N-1-3} \\
  %      X_{3,1} & \cdots &  X_{3,N-1-3}
  %  \end{bmatrix}.
%\end{equation*}

\subsection{Linear constraint based on LRPV1}

%Proposed LRPV1 needs the precondition that $M>N+3$.
From Eqs. (\ref{thrm1}) and (\ref{eq9}), we split $T_1^\ast$ as $T_{1}^\ast =\begin{bmatrix}
      T_{11}^\ast & T_{12}^\ast
\end{bmatrix}$, where $T_{11}^\ast\in \mathbb{R}^{(M-1) \times (N-1+3)}$ and $T_{12}^\ast \in \mathbb{R}^{(M-1) \times (N-1-3)}$, then we have 
\begin{equation}
\label{eq18}
rank(\begin{bmatrix}
    T_{11}^\ast & T_{12}^\ast
\end{bmatrix}=rank(T_{11}^\ast) \leq N-1+3.
\end{equation}

From Eq. (\ref{eq18}), we can assume there is a matrix \(Z\) that enables  
\begin{equation}
\label{eq17}
        T_{11}^\ast Z=T_{12}^\ast,
\end{equation}
where $Z\in \mathbb{R}^{(N-1+3) \times (N-1-3)}$ is unknown and is estimated in Section VI-E.

%In addition, it can seen from \textit{Appendix A} that the matrix $Z$ is constructed by both $X$ in equ (\ref{eq15}) and identity matrix, i.e.,
%\begin{equation*}
  %  Z=\begin{bmatrix}
   %   X \\ -I \\X
  %  \end{bmatrix}=\begin{bmatrix}
  %       X_{1,1} &  X_{1,2} & \cdots  & X_{1,N-1-3} \\
  %       X_{2,1} &  X_{2,2} & \cdots  & X_{2,N-1-3} \\
  %        X_{3,1} &  X_{3,2} & \cdots & X_{3,N-1-3}\\
 %         -1 & 0 & \cdots & 0\\
 %         0 & -1 & \cdots & 0 \\
  %        \vdots  & \vdots & \ddots & \vdots   \\
   %       0 & 0 & \cdots & -1 \\
   %        X_{1,1} &  X_{1,2} & \cdots  & X_{1,N-1-3} \\
   %      X_{2,1} &  X_{2,2} & \cdots  & X_{2,N-1-3} \\
   %       X_{3,1} &  X_{3,2} & \cdots & X_{3,N-1-3}\\
   %   \end{bmatrix},
%\end{equation*}
%in this way, there is no additional unknown variables need to be estimated in comparison with the equ (\ref{eq15}). 

\subsection{Linear constraint based on LRPV2}

%Proposed LRPV2 needs the precondition that $N>M+3$. 
From Eqs. (\ref{thrm2}) and (\ref{eq11}), we can split $T_2^\ast$ as $T_{2}^\ast =\begin{bmatrix}
    T_{21}^\ast & T_{22}^\ast
\end{bmatrix}$ where $T_{21}^\ast \in \mathbb{R}^{(N-1) \times (M-1+3)}$ and $T_{22}^\ast \in \mathbb{R}^{(N-1) \times (M-1-3)}$. Then we have 
\begin{equation}
\label{eq20}
rank(\begin{bmatrix}
    T_{21}^\ast & T_{22}^\ast
\end{bmatrix})=rank(T_{21}^\ast) \leq M-1+3.
\end{equation}

From Eq. (\ref{eq20}), we can assume there is a matrix \(W\) that enables  
\begin{equation}
\label{eq19}
        T_{21}^\ast W=T_{22}^\ast,
\end{equation}
where $W\in \mathbb{R}^{(M-1+3) \times (M-1-3)}$ is unknown and is estimated in Section VI-E.

%In addition, it can seen from \textit{Appendix B} that the matrix $W$ is constructed by both $\hat{X}$ in (\ref{newmain2}) and identity matrix, i.e.,
%\begin{equation*}
 %   W=\begin{bmatrix}
 %     \hat{X} \\ -I \\ \hat{X}
 %   \end{bmatrix}=\begin{bmatrix}
 %         \hat{X}_{1,1} &  \hat{X}_{1,2} & \cdots  &  \hat{X}_{1,M-1-3} \\
   %       \hat{X}_{2,1} &   \hat{X}_{2,2} & \cdots  &  \hat{X}_{2,M-1-3} \\
    %       \hat{X}_{3,1} &   \hat{X}_{3,2} & \cdots &  %\hat{X}_{3,M-1-3}\\
   %       -1 & 0 & \cdots & 0\\
   %       0 & -1 & \cdots & 0 \\
  %        \vdots  & \vdots & \ddots & \vdots   \\
  %        0 & 0 & \cdots & -1 \\
 %          \hat{X}_{1,1} &  \hat{X}_{1,2} & \cdots  & \hat{X}_{1,M-1-3} \\
  %       \hat{X}_{2,1} & \hat{X}_{2,2} & \cdots  & \hat{X}_{2,M-1-3} \\
  %       \hat{X}_{3,1} &  \hat{X}_{3,2} & \cdots & \hat{X}_{3,M-1-3}\\
  %    \end{bmatrix},
%\end{equation*}
%in this way, there is no additional unknown variables need to be estimated in comparison with the (\ref{newmain2}). 
\subsection{Linear constraint based on LRPV3}

From Eqs. (\ref{thrm3}) and (\ref{eq13}), we define $M_N$ as $min(N-1+3, M-1+3)$ and split $T_3^\ast$ as $T_3^\ast = \begin{bmatrix}
    T_{31}^\ast & T_{32}^\ast
\end{bmatrix}$ where $T_{31}^\ast \in \mathbb{R}^{2(M-1) \times M_N}$ and $T_{32}^\ast\in \mathbb{R}^{2(M-1) \times(2(N-1)-M_N)}$.
 Then we have 
\begin{equation}
\label{eq22}
rank(\begin{bmatrix}
    T_{31}^\ast & T_{32}^\ast
\end{bmatrix})=rank(T_{31}^\ast) \leq M_N.
\end{equation}

From Eq. (\ref{eq22}), we can assume there is a matrix \(Y\) that enables  
\begin{equation}
\label{eq21}
        T_{31}^\ast Y=T_{32}^\ast,
\end{equation}
where $Y\in \mathbb{R}^{M_N \times (2(N-1)-M_N)}$ is unknown and is estimated in Section VI-E.

\subsection{Algorithm}
\begin{figure}[!t]
   \centering
    \includegraphics[trim=0.2cm 0.3cm 0.2cm 0.2cm, clip=true, width=0.3\textwidth ]{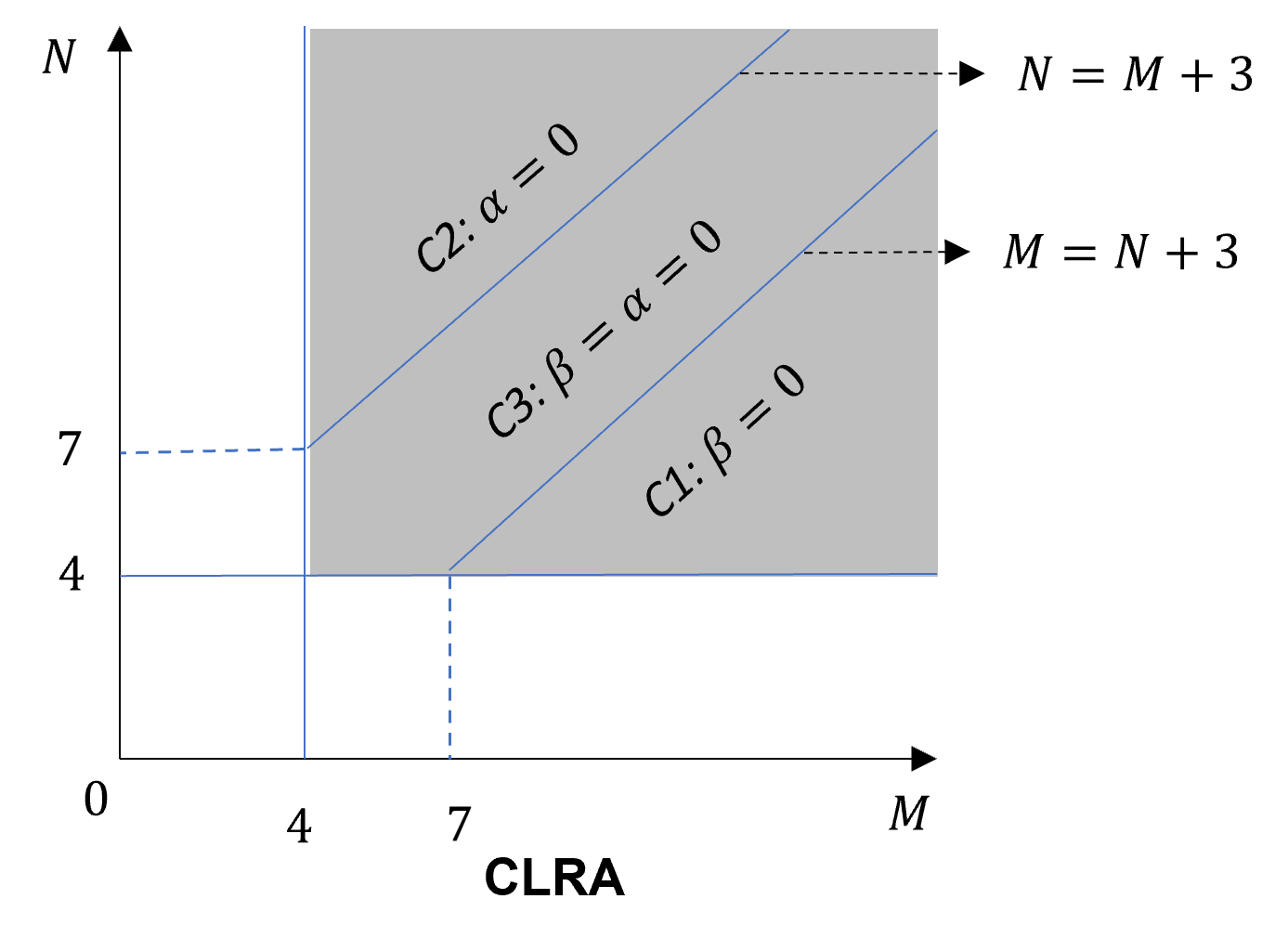}
    \caption{The illustration for combinations of four low-rank properties of proposed CLRA method with different number of  microphones \(M\)  and sources \(N\) ($\alpha$ for LRPV1, $\beta$ for LRPV2, $\gamma$ for LRPV3; $\alpha=\beta=\gamma=0$ in the shadow area for STLS~\cite{8}).}
    \label{figure:precodition}
    %\redcolor{add x and y axis [0, 20]}}
\end{figure}
%We consider two different situations for designing the algorithms: 1) $M\geq N$; 2) $M<N$.  

%1) If $M\geq N$, based on 
The STLS \cite{8,39} utilizes the LRP only for UTIm estimation, so that the solution of UTIm is easy to stuck into local minimal value,  this limits both the recovery and convergency rate for UTIm estimation, and  the estimation errors of UTIm tends to be large in noise environments. To find global optimal solution for UTIm, three variants of LRP are proposed in Section V, and based on the linear constraints formulated from those four low-rank properties, LRP, LRPV1, LRPV2 and LRPV3, %(LRP, LRPV1, LRPV2 and LRPV3) in (\ref{eq15}), (\ref{eq17}), (\ref{eq19}) and (\ref{eq21}), 
we have the objective function:
%
%We define the objective function based on (\ref{eq15}), (\ref{eq17}), (\ref{eq19}) and (\ref{eq21}):  
\begin{align}
\label{eq23}
      f(\delta, \eta, X, Y, Z, W)   = & \|U\|_F^2  +\lambda^2 \|(A+F)X-(B+G)\|_F^2 \nonumber \\ 
     & +\alpha^2\| T_{11}^\ast Z-T_{12}^\ast\|_F^2 \nonumber\\
     & + \beta^2\| T_{21}^\ast W-T_{22}^\ast\|_F^2  \nonumber\\
      & +\gamma^2 \| T_{31}^\ast Y-T_{32}^\ast\|_F^2 ,
\end{align}
% \end{equation*}
% \begin{equation}
%     \label{eq23}
%              +\gamma^2 \| T_{31}^\ast Y-T_{32}^\ast\|_F^2 +\alpha^2\| T_{11}^\ast Z-T_{12}^\ast\|_F^2
%               + \beta^2\| T_{21}^\ast W-T_{22}^\ast\|_F^2,
% \end{equation}
 %   
%2) If $M< N$, based on (\ref{newmain2}), (\ref{eq19}) and (\ref{eq21_1}), we can have the objective function, %
%\begin{equation*}
 %   min f(\delta, \eta, \hat{X})=\|U\|_F^2 +\lambda^2 \|(\hat{A}+\hat{F})\hat{X}-(\hat{B}+\hat{G})\|_F^2 
 %   \end{equation*}
  %  \begin{equation}
  %  \label{eq23_1}
  %           +\gamma^2 \| {\hat{T}_{31}^\ast}^T \hat{Y}-{\hat{T}_{32}^{\ast}}^T\|_F^2 +\beta^2\| T_{21}^\ast W-T_{22}^\ast\|_F^2
       %       + \beta^2\| T_{21}^\ast W-T_{22}^\ast\|_F^2
  %  \end{equation}
where $\| \bullet \|_F$ is Frobenius norm, $\| U \|_F^2$ is regularization term~\cite{8,39}, and \(\lambda\),  \(\alpha\),  \(\beta\) and \(\gamma\) are penalty parameters for the four low-rank properties, respectively (\(\lambda\) for LRP; \(\alpha\) for LRPV1; \(\beta\) for LRPV2; \(\gamma\) for LRPV3). Upon inspection of Eq. (\ref{eq23}), We can see that when \(\gamma=\alpha=\beta=0\), proposed CLRA method degenerates to STLS that uses the LRP only for UTIm estimation~\cite{8,39}. 

In addition, for any given number of microphones \(M\) and any given number of sources \(N\), we can distinguish three cases, i.e., 1) Case 1 (C1): $M-N>3$; 2) Case 2 (C2): $N-M>3$; 3) Case 3 (C3): $|M-N|\leq 3$.  From $M-1>N-1+3$ in Eq. (\ref{eq9}) and $N-1>M-1+3$ in Eq. (\ref{eq11}), we can see that LRPV1 and LRPV2 have low-rank properties only in C1 and C2, respectively, then we can summarize the combination way for four low-rank properties  as

   1) $\beta=0$ in C1;
   
    2) $\alpha=0$ in C2; and
    
    3) $\alpha=0$ and $\beta=0$ in C3.  
 Fig.~\ref{figure:precodition} illustrates the combination way of four low-rank properties for CLRA.
%
% \par
% \begin{figure}[!ht]
%     \centering
%     \includegraphics[width=0.4\textwidth]{Fig. 1.png}
%     \caption{The description for precondition of proposed CLRA method with different number of  microphones \(M\)  and sources \(N\).}
% \end{figure}
%
%\begin{figure}[ht]
% \begin{center}
%\includegraphics[width=1\linewidth]{mn.png}
%\caption{The description for precondition of STLS and proposed CLRA method with different number of microphones \(M\)  and sources \(N\).}
 %\end{center}
% \redcolor{add x and y axis [0, 20]}
 %\end{figure}
 
Next, we present a general and complete solution for the minimization of above objective function. We use the column-wise matrix vectorization, $v(\cdot)$ (i.e., $v(X)=\begin{matrix}
    [X_{1,1}, \cdots, X_{3,1}, X_{1,2}, \cdots, X_{3,2}, \cdots, X_{1,(N-1-3)},\end{matrix}$ $ \cdots, X_{3,(N-1-3)}]^T$), to define  
\begin{equation*}
p= \begin{bmatrix}
  \dot{\delta}^T & \dot{\eta}^T & v(X)^T & v(Y)^T & v(Z)^T & v(W)^T 
\end{bmatrix}^T,
\end{equation*}
and 
\begin{equation*}
    q=\begin{bmatrix}
f_A^T & \lambda f_B^T & \gamma f_C^T & \alpha f_D^T & \beta f_E^T
\end{bmatrix}^T,
\end{equation*}
%$q=\begin{bmatrix}
%f_A^T & \lambda f_B^T & \gamma f_C^T & \alpha f_D^T & \beta %f_E^T
%\end{bmatrix}^T$
where
\begin{equation}
 \label{eq29}
 \begin{cases}
 f_A= v(U) \\
 f_B= v((A +F)X-B-G) \\
 f_C= v(T_{31}^\ast Y-T_{32}^\ast) \\
 f_D= v(T_{11}^\ast Z-T_{12}^\ast) \\
 f_E= v(T_{21}^\ast W-T_{22}^\ast) 
 \end{cases},
%q = 
%\begin{bmatrix}
%v(U) \\
%\lambda v((A +F)X-B-G) \\
%\gamma v(T_{31}^\ast Y-T_{32}^\ast) \\
%\alpha v(T_{11}^\ast Z-T_{12}^\ast) \\
%\beta v(T_{21}^\ast W-T_{22}^\ast)
%\end{bmatrix}
 \end{equation}
and then re-write the objective function Eq. (\ref{eq23}) as $f\left(p\right)= \| q \|^2_2$,
% where \(x\), \(y\), \(z\) and \(w\) are defined as
% \begin{equation*}
%     x={vec}_c(X)=[
%         X_{1,1} & \cdots & X_{3,1} & \cdots
% \end{equation*}
% \begin{equation*}
%  \label{eq24}
%     \begin{matrix}
%   X_{1,(N-1-3)} & \cdots & X_{3,(N-1-3)}
%     \end{matrix}]^T,
% \end{equation*}
% \begin{equation*}
%     y={vec}_{c}(Y)=[
%     \begin{matrix}
%      Y_{1,1} & \cdots &  Y_{M_{N},1} &  \cdots
%     \end{matrix}
% \end{equation*}
% \begin{equation*}
%  \label{eq25}
%     \begin{matrix}
%   Y_{1,2(N-1)-M_{N}} & \cdots & Y_{M_{N},2(N-1)-M_{N}}
%     \end{matrix}
%     ]^T,
% \end{equation*}
% \begin{equation*}
%      z={vec}_{c}(Z)=[
%     \begin{matrix}
%      Z_{1,1} & \cdots &  Z_{N-1-3,1}  & \cdots
%     \end{matrix}
% \end{equation*}
% \begin{equation*}
%  \label{eq26}
%     \begin{matrix}
%   Z_{1,N-1+3} & \cdots & Z_{N-1-3, N-1+3}
%     \end{matrix}]^T,
% \end{equation*}
% and 
% \begin{equation*}
%     w={vec}_{c}(W)=[
%     \begin{matrix}
%      W_{1,1} & \cdots &   W_{M-1-3,1}  &  \cdots
%     \end{matrix}
% \end{equation*}
% \begin{equation}
%  \label{eq27}
%     \begin{matrix}
%      W_{1,M-1+3} & \cdots & W_{M-1-3, M-1+3}
%     \end{matrix}]^T,
% \end{equation}
% where ${vec}_{c}$ denotes the operation from matrix to column vector. 
% \begin{equation*}
% \label{eq28}
%  {f\left(p\right)}= \sum_{q=1}^{Q} \hat{f}_q^2,
% \end{equation*}
% the rightmost term \(f=\begin{bmatrix}
%   f_A^T, & \lambda f _B^T, & \gamma f_C^T, & \alpha f_D^T, & \beta f_E^T
% \end{bmatrix}^T  =[f_1,f_2,\cdots,f_Q]^T\) with five columns, 
where the dimension of the vector $q$ and vector $p$ are  $Q=(M-1)(8(N-1)-2M_N-6)-3(N-1)$ and $P=M+N-1+3(N-1-3)+M_N(2(N-1)-M_N)+(N-1+3)(N-1-3)+(M-1+3)(M-1-3)$, respectively.

\begin{figure}[!t]
   \centering
\includegraphics[width=0.45\textwidth ]{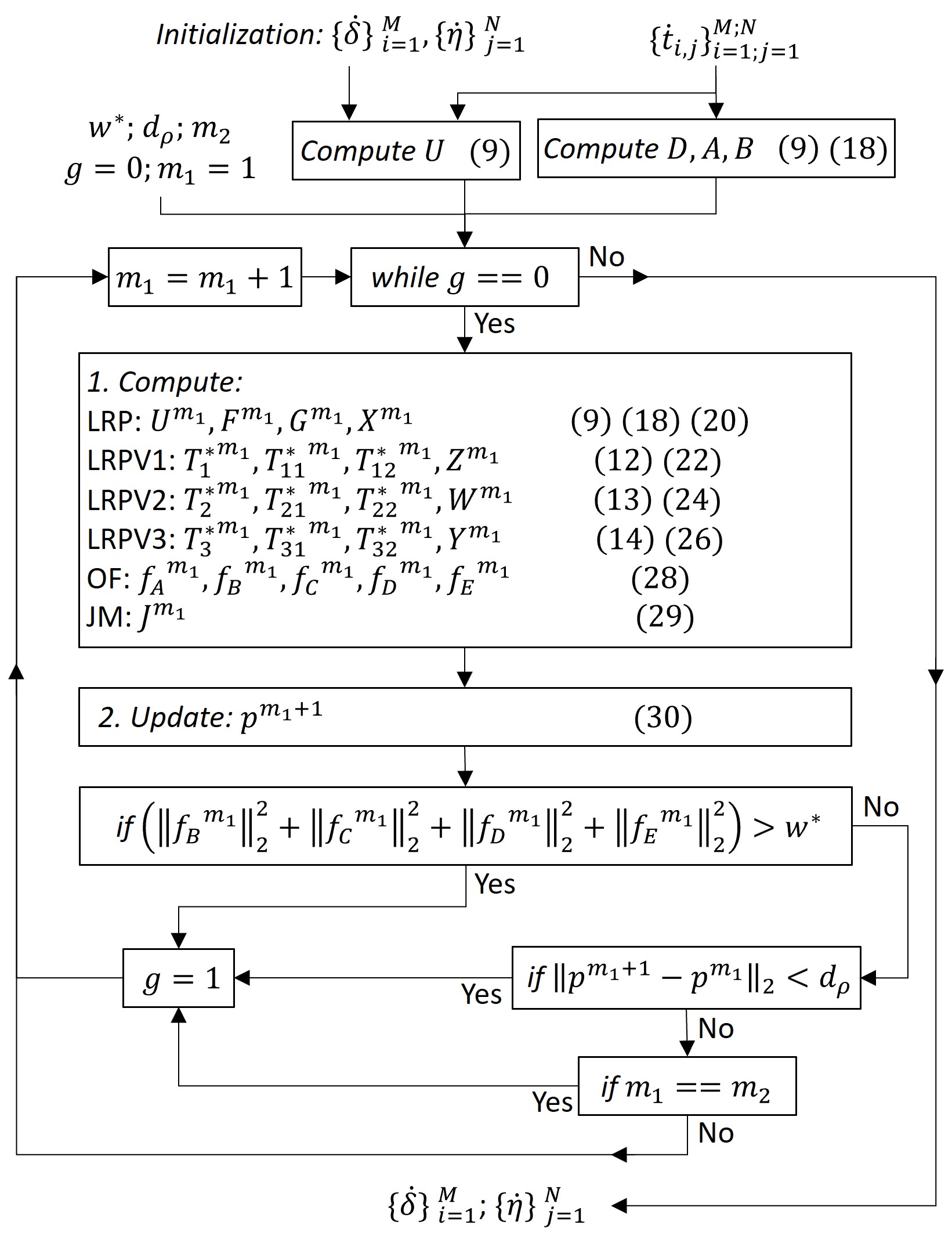}

    \caption{The flowchart for proposed CLRA method ($w^\ast$: threshold for divergence; $d_p$: stopping threshold for iterations; $m_2$: maximal number of iterations; $m_1$: $m_1^{th}$ iteration; OF: Objective function; JM: Jacobian matrix; $\|\bullet\|_2$: 2 norm).}
    \label{Algorithm}
    %\redcolor{add x and y axis [0, 20]}}

\end{figure}

Finally, we solve the nonlinear least square problem by minimizing $f(p)$ with the Gauss-Newton algorithm~\cite{40}.  Therefore we compute the Jacobian matrix
\begin{equation}
    J=\partial q/\partial p \in \mathbb{R}^{Q \times P},
\end{equation}
which is the derivation of vector $q$ with respect to vector $p$ (see details in \textit{Section A4 of Supplementary Material}), and we update \(p\) in an iterative way as
\begin{equation}
\label{eq31}
    p^{m+1}=p^{m}-{{(J^{m}}^TJ^{m})}^{-1}{J^{m}}^Tq^{m} ,
\end{equation}
where \(m\) denotes $m^{th}$ iteration and the general flowchart of proposed CLRA can be seen from Fig. \ref{Algorithm}.

% , and the pseudo-code of proposed CLRA can be seen at \textit{Algorithm}. 
%Also, we will name the proposed CLRA method as CLRA1 when $|M-N|\leq 3$; if $M>N+3$, the proposed CLRA method will be CLRA2; and the proposed CLRA method will be CLRA3 if $N>M+3$.

\begin{figure*}[!t]
\begin{center}
{\includegraphics[trim=5.0cm 6cm 1.5cm 6cm, clip=true,width=0.9\linewidth]{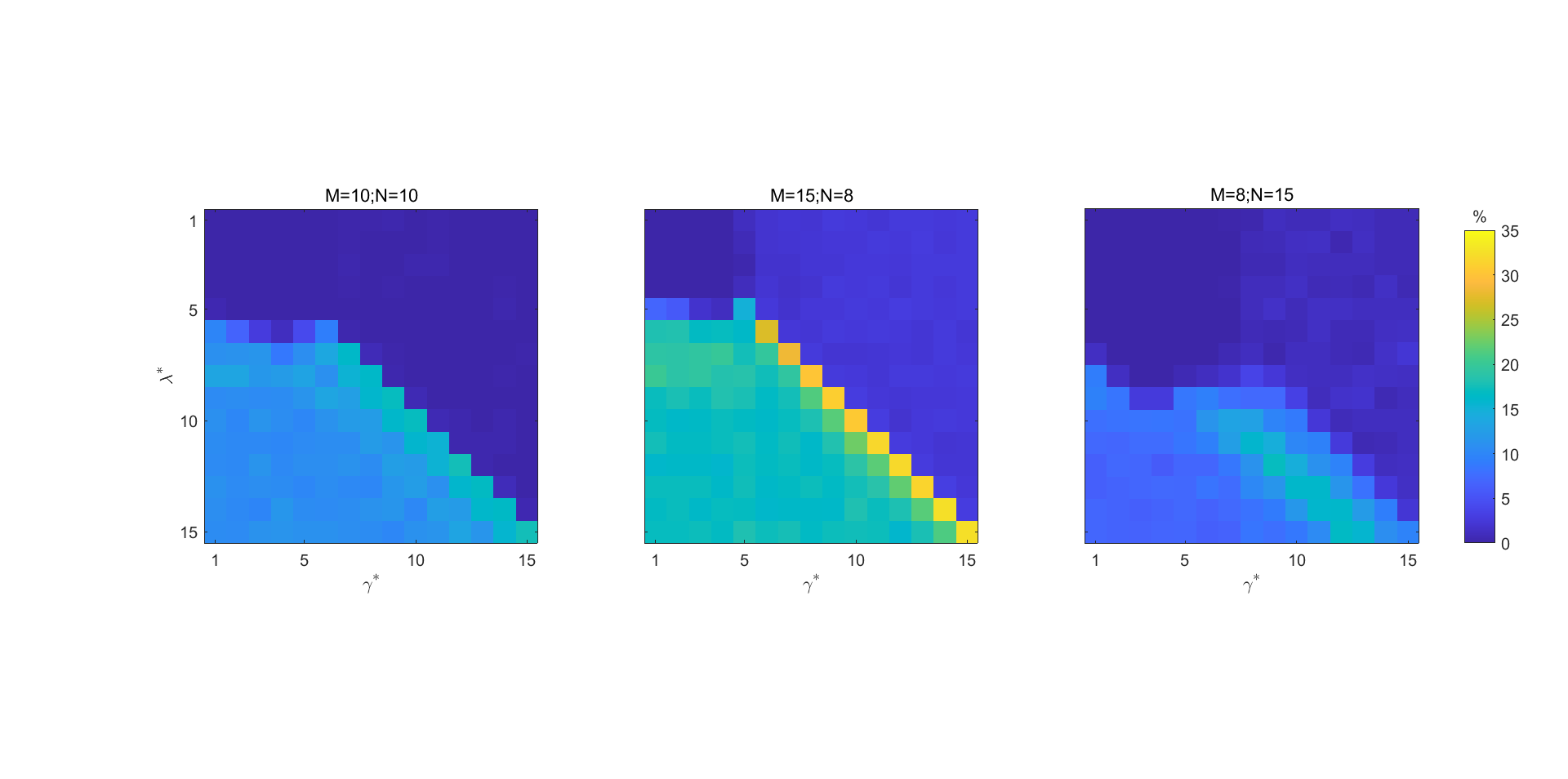}}
\caption{The parameters impact ($\lambda^\ast$ for LRP and $\gamma^\ast$ for LRPV3) on proposed CLAR1 in terms of recovery rate with three different cases.}
\label{figureCLRA1}
\end{center}
\end{figure*}
%\begin{figure*}[ht]
%\centering
%\subfigure[C3]{\includegraphics[trim=0.2cm 0cm 0.7cm 0.2cm, clip=false, width=0.3\linewidth]{pa1010_1.png}}
%\subfigure[C1]{\includegraphics[trim=0.2cm 0cm 0.7cm 0.2cm, clip=false,width=0.3\linewidth]{pa158_1.png}}
%\subfigure[C2]{\includegraphics[trim=0.2cm 0cm 0.7cm 0.2cm, clip=false,width=0.3\linewidth]{pa815_1.png}}
%\caption{The parameters impact ($\lambda^\ast$ for LRP and $\gamma^\ast$ for LRPV3) on proposed CLAR1 in terms of recovery rate with three different situations.}
%\label{figureCLRA1}
%\vspace{-1em}
%\end{figure*}
\section{Experimental Results}
 The simulation settings are discussed in Subsection A and the impact of parameters of proposed three variants of LRP on CLRA method is analyzed in Subsection B. Next, Subsections C, D and E show the performance of the proposed CLRA method in comparison with the STLS~\cite{8,39} and auxiliary function-based algorithms~\cite{33}. Then, we conduct robust analysis in Subsection F for both STLS and CLRA by adding noise to both simulation and real data. Finally, we show the limitations of proposed  variants of LRP and CLRA method in Subsection G. 
 \vspace{-1em}
\subsection{Simulation Setup}
 The simulation data for location of microphones and sources, microphones start time and sources emission time  is generated randomly by MATLAB R2019a on a computer with 3.7-GHz CPU,
six cores, and 16.0G RAM. In more details, the location of microphones and sources are randomly generated with uniform distribution from the ranges  $10m \times 10m \times 3m$ which is suitable for the real applications \cite{34}. In addition, the start time $(\delta)$ and emission time $(\eta)$ are randomly generated with uniform distribution from range $\begin{bmatrix}
    -1 & 1
\end{bmatrix}s$~\cite{34} and we set the speed of sound $c=340 m/s$. Besides, for the parameters of proposed CLRA (see Fig. \ref{Algorithm}), i.e., $w^\ast$, $d_p$ and $m_2$, we set $w^\ast=10^{30}$, $d_p=10^{-9}$ and $m_2=100$, it implies that if the values of objective function is larger than $10^{30}$, or the difference of values for variable $p$ in Eq. (\ref{eq29}) between two adjacent iterations is less than $10^{-9}$, or the iterative number is larger than 100, we end the process of CLRA.

%Usually, the term $\frac{\|r_i-s_j\|}{c}$ (the propagation time of audio signal emitted from $j^{th}$ source to $i^{th}$ microphone) in (\ref{eq1}) is quite small in comparison with the capture time ($\delta_i$), onset time ($\eta_j$) and the time of arrival ($t_{i,j}$). It leads to the unknown timing information (capture time $\delta_i$ and onset time $\eta_j$), known time of arrival $t_{i,j}$ and the unknown propagation time of audio signal emitted from $j^{th}$ source to $i^{th}$ microphone on the different order of magnitude. Thus, we provide a pre-processing step for processing the known $t_{i,j}$, this pre-processing step could let the values of unknown timing information (capture time $\delta_i$ and onset time $\eta_j$), known time of arrival ($t_{i,j}$) and the unknown sound propagation time from $j^{th}$ source to $i^{th}$ microphone in the same order of magnitude, more importantly, this pre-processing step could show the boundary of known TOA and unknown timing information, for the details, it can be found in \textit{Appendix E}. Besides, we also proved that the TDOA is same as the TOA, and the time offset from TDOA is also same as the unknown timing information of TOA, for more details, please see \textit{Appendix F}.

Multiple initializations are set for each configuration to showcase the CLRA method's ability to resist changes of initialization. In a specific configuration, only one initialization for the microphones start time and sources emission time is required. However, in some cases, the random initialization cannot attain the globally optimal solution of UTIm. Therefore, by calculating the number of global optimal solutions of UTIm achieved by the proposed CLRA method in multiple random initializations and then comparing it with the number of globally optimal solution of UTIm achieved by other state-of-the-art methods, the excellent performance of recovery rate achieved by the proposed CLRA method is demonstrated. The definition of the recovery rate is 
\begin{equation}
\label{eq33}
    Rr(M,N)=\frac{\sum_{i=1}^{Nc(M,N)} Ne_i(M,N)}{I_n(M,N)Nc(M,N)},
\end{equation}
where \(Ne_i(M,N)\) is the number of globally optimal solutions of the initializations for $i^{th}$ configuration, \(Nc(M,N)\) is the total number of configurations, and $I_n(M,N)$ is the total number of initializations for each configuration. Usually, the sampling rate of microphones is less than 100k Hz, and for one audio signal, the difference between two adjacent samples is about the magnitude of $10^{-5}s$ (i.e., $\frac{1s}{100000}=10^{-5}s$), thus there are about the magnitude of $10^{-5}s$ errors for TOA/TDOA  by using generalized cross-correlation with phase
transform (GCC-PHAT)~\cite{gccp} method to estimate TOA/TDOA with audio signals. For example, if the sampling rate of mics is 44.1k Hz and the accurate sampling point of TOA is 810.4356, GCC-PHAT will choose 811 as the sampling point that could achieve the maximal cross-correlation function value, this could introduce $1.2798\times 10^{-5}s$ errors for TOA/TDOA. Based on this principle, if the difference between the estimation value of UTIm and the ground truth of UTIm is less than $10^{-4}s$, we assume the UTIm to be recovered successfully, and $er$ is defined as
\begin{equation}
\label{er}
    er=\frac{\sum_{i=1}^{M} \|\dot{\delta}_i - \hat{\dot{\delta}}_i\|}{M}+ \frac{\sum_{j=1}^{N} \|\dot{\eta}_j - \hat{\dot{\eta}}_j\|}{N}
\end{equation}
where $\dot{\delta}_i$ and  $\dot{\eta}_j$ are the ground truth of UTIm, and $\hat{\dot{\delta}}_i$ and  $\hat{\dot{\eta}}_j$ are the estimation value of UTIm.

In addition, the convergency rate is defined as
\begin{align}
  \label{eq34}
  & Cr(M,N)=\frac{Ce(M,N)}{Nc(M,N)}, \nonumber \\
  &  Ce(M,N)=\sum_{i=1}^{Nc(M,N)}sr_i, \nonumber \\
  &   {sr}_i=\Bigg \{
    \begin{matrix}
    1   & Ne_i(M,N)\neq 0 \\
    0   & Ne_i(M,N)= 0
    \end{matrix},
\end{align}
%\begin{equation*}
%\label{eq34}
 %   Cr(M,N)=\frac{Ce(M,N)}{Nc(M,N)} 100 \%,
%\end{equation*}
%\begin{equation*}
%\label{eq35}
 %  Ce(M,N)=\sum \uparrow_{sr_i,0},
%\end{equation*}
%\begin{equation}
%\label{eq36}
 %   \uparrow_{sr_i,0}=\Bigg \{
 %   \begin{matrix}
%    1   & sr_i=0 \\
 %   0   & sr_i \neq 0
 %   \end{matrix},
%\end{equation}
where \(Cr(M,N)\) and \(Ce(M,N)\) are convergency rate and the total number of successful recovery in all configurations, respectively, \(sr_i\) is the successful recovery for $i^{th}$ configuration, i.e. if there is one successful recovery for all initializations in one configuration, this configuration is regarded as successful recovery. As suggested by the name, the convergency rate means that the ratio of number for successful recovery configurations to total number of configurations and it can reach near 100 \% when the number of microphones and/or sources is sufficient, and the main aim of this metric is to show the ability of proposed CLRA to resist the change of configurations when the number of microphones or sources is not sufficient.

\subsection{Parameters Analysis}

% \bluecolor{in order to make a clear analysis, which means that the contribution of proposed each rank property will be shown, we will name proposed CLRA method as different names when some parameters have been set to be zero (see Fig. 1(b)). , and we will name the proposed CLRA method as CLRA1 when both \(\alpha\) and \(\beta\) set to be zero; also, if \(\beta\) and \(\alpha\) equal to 0, the proposed CLRA will name as CLRA2 and CLRA3, respectively, i.e. \(CLRA:\big\{\begin{matrix} CLRA1 & \alpha=0, \beta=0;   \\ CLRA2 &  \beta=0; \\ CLRA3 & \alpha=0; \end{matrix}\)},}

This subsection analyzes the parameters for proposed three variants of LRP in CLRA. 
%this is because that proposed matrix \(T_3^\ast\) of LRPV3 in (\ref{eq12}) just have one precondition, i.e., $M> 3+1$ and $N> 3+1$. And the proposed matrices  \(T_1^\ast\) of LRPV1 in (\ref{eq8}) and \(T_2^\ast\) of LRPV2 in (\ref{eq10}) have one more precondition (\(M>N+3\) for \(T_1^\ast\) and \(N>M+3\) for \(T_2^\ast\)).
We name the proposed CLRA as: 1) CLRA1 if $\alpha=\beta=0$ (combination of LRP and proposed LRPV3); 2) CLRA2 if $\alpha=\gamma=0$ (combination of LRP and proposed LRPV2); 3) CLRA3 if $\beta=\gamma=0$ (combination of LRP and proposed LRPV1).
%\(CLRA:\big\{\begin{matrix}CLRA1 & \alpha=0, \beta=0;   \\ CLRA2 &  \beta=0; \\ CLRA3 & \alpha=0; \end{matrix}\)};
 We set the number of configurations $N_c(M,N)=10$  and the number of initializations $I_n(M,N)=100$ for each configuration, thus, there are 1000 implementations for fixed $M$ and $N$, and the impact of the parameters on proposed CLRA1, CLRA2 and CLRA3 can be shown. 
  
In addition, as can be seen from Fig. \ref{figure:precodition}, the parameter \(\beta\) and \(\alpha\) of the proposed CLRA method are zero in C1 and C2, respectively; And in C3, both the  \(\alpha\) and  \(\beta\) are zero. However, when both \(\alpha\) and \(\beta\) are  zero, the proposed CLRA also works different number of microphones \(M\) and sources \(N\)  since proposed LRPV3 always has low-rank property among C1, C2 and C3.
 Thus, the parameters of proposed CLRA1 (\(\lambda\) and \(\gamma\) in Eq. (\ref{eq23})) are analyzed in C1, C2 and C3, and we set $M=15$, $N=8$ for C1, $M=8$, $N=15$ for C2 and $M=10$, $N=10$ for C3. After these two parameters \(\lambda\) and \(\gamma\) are analysed, we analyse the parameter \(\beta\) of CLRA2 with \(M=15\) and \(N=8\) while  \(M=8\) and \(N=15\) are fixed for parameter \(\alpha\) of CLRA3.

 \pgfplotstableread{material/parameters_analysis.txt}\paramsanalysis
\begin{figure}[!t]
    \centering
    \begin{tikzpicture}
    \begin{axis}[
        width=.90\columnwidth,
        height=.5\columnwidth,
        ymin=0,ymax=40,
        ylabel={Recovery rate (\%)},
        tick label style={font=\footnotesize},
        xmin=0, xmax=15.5,
        xlabel={$\alpha^\ast,\beta^\ast$},
        label style={font=\footnotesize},
    ]
    \addplot+[solid, color=blue, mark=none] table[x=x, y=alpha]{\paramsanalysis};
    \addplot+[solid, color=red, mark=none] table[x=x, y=beta]{\paramsanalysis};
    \end{axis}
    \end{tikzpicture}
    \caption{
 The effect of $\beta^\ast$ on CLRA2 with $M=8$, $N=15$ (\protect\raisebox{1pt}{\protect\tikz \protect\draw[red,fill=red] (0,0) rectangle (1.ex,1.ex);}) and $\alpha^\ast$ on CLRA3 with $M=15$, $N=8$ (
    \protect\raisebox{1pt}{\protect\tikz \protect\draw[blue,fill=blue] (0,0) rectangle (1.ex,1.ex);}).}
    \label{fig:analysislayers}
\end{figure}
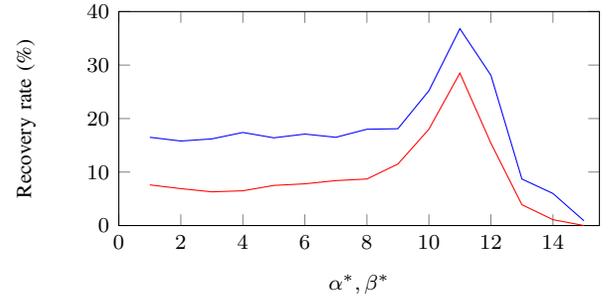
\begin{figure*}[t]
\begin{center}
{\includegraphics[trim=4.0cm 8.4cm 1.5cm 8.4cm, clip=true,width=0.9\linewidth]{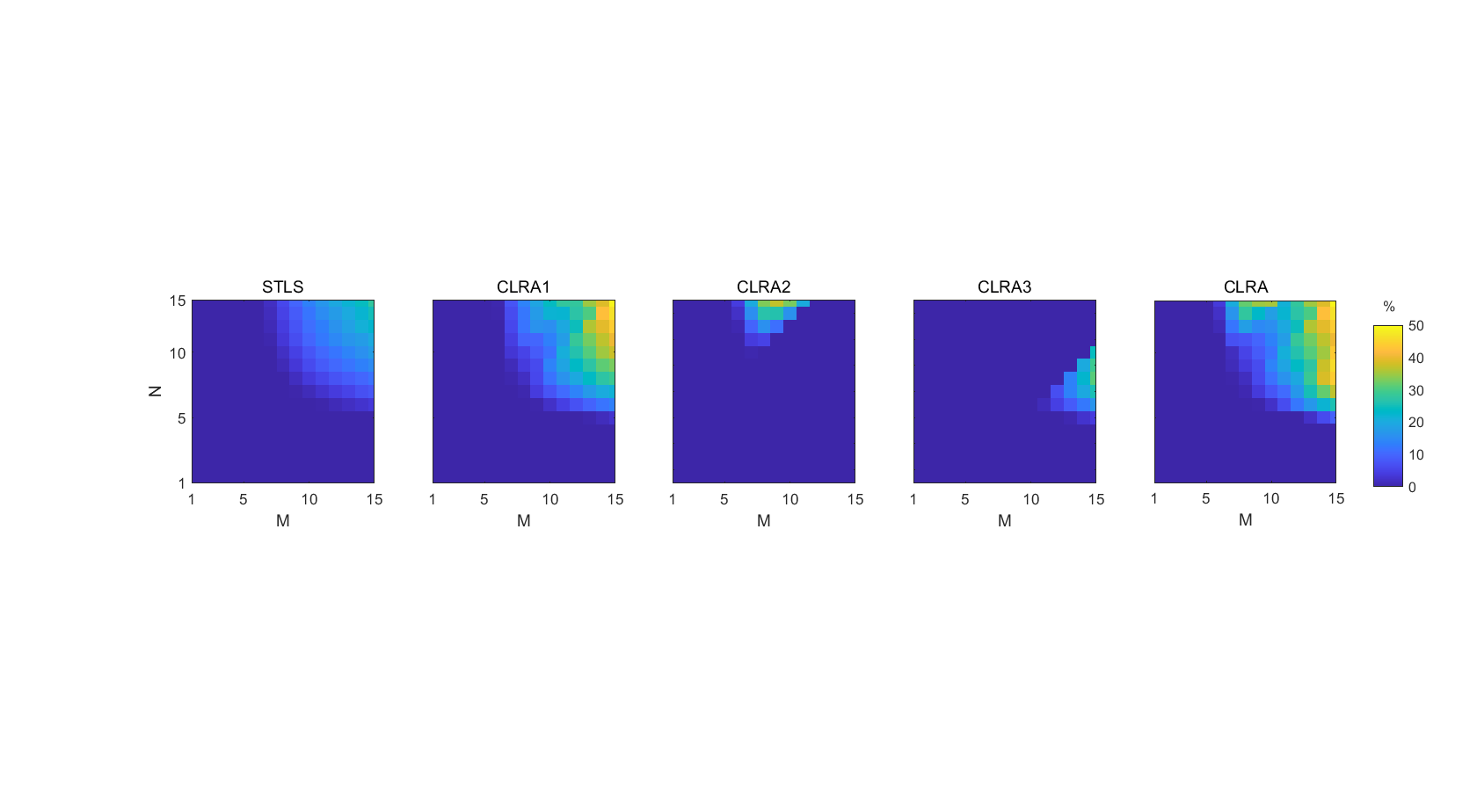}}
\caption{The performance comparison for STLS and proposed CLRA (CLAR1: \(\alpha=\beta=0\); CLAR2: \(\alpha=\gamma=0\); CLAR3:$\beta=\gamma=0$) in terms of recovery rate.}
\label{figurecompari}
\end{center}
\end{figure*}

\begin{figure*}[!t]

\centerline{\subfigure[]{\includegraphics[trim=4.0cm 8.4cm 1.3cm 8.4cm, clip,width=0.9\linewidth]{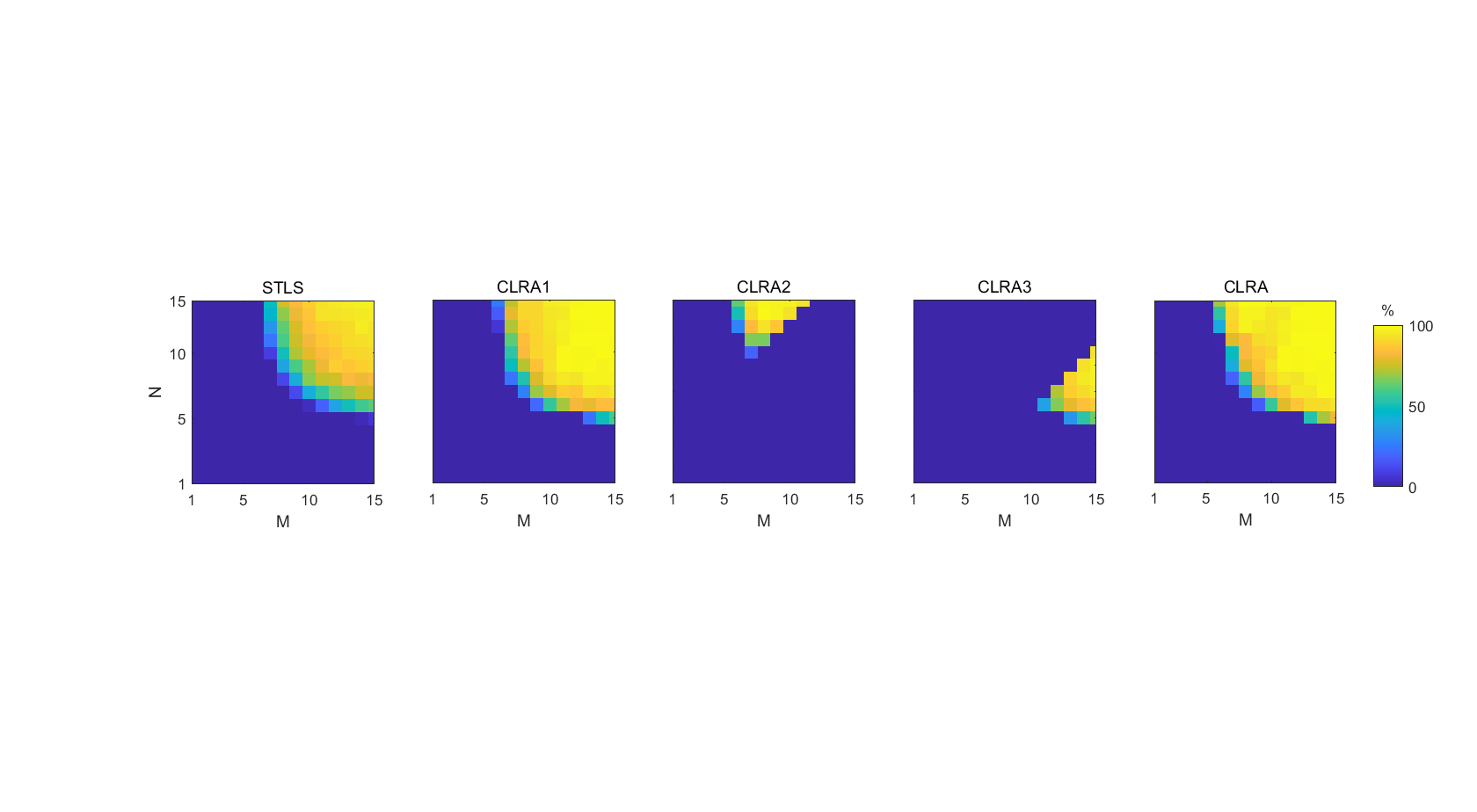}}}
\centerline{\subfigure[]{\includegraphics[trim=4.0cm 8.4cm 1.3cm 8.4cm, clip,width=0.9\linewidth]{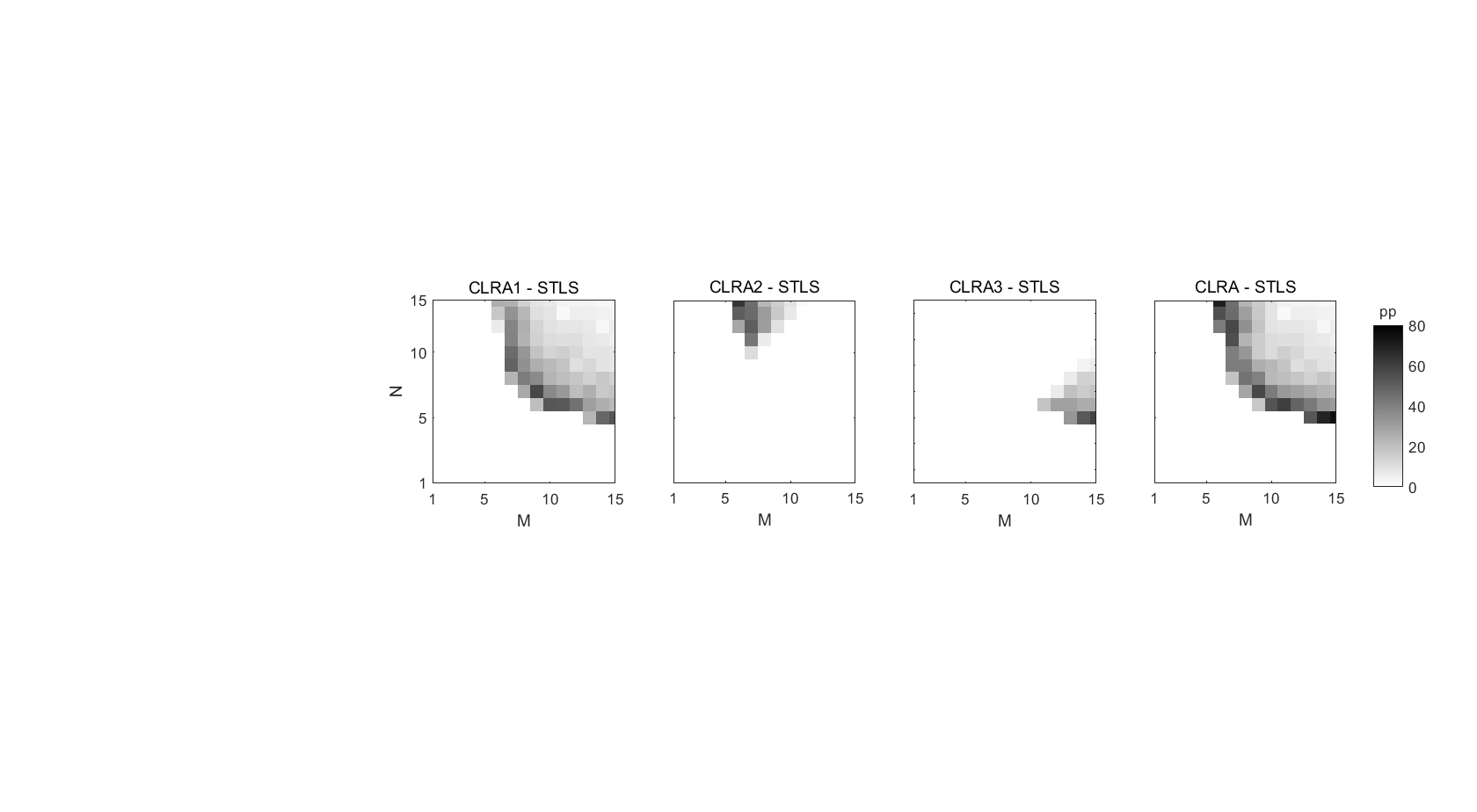}}}

\caption{The performance comparison for STLS and proposed CLRA in terms of (a) convergency rate; and (b) percentage point of convergency rate, where CLAR1: \(\alpha=\beta=0\), CLAR2: \(\alpha=\gamma=0\), CLAR3: $\beta=\gamma=0$,  pp: percentage point.}
\label{figurecompari1}

\end{figure*}

\textit{1) Parameters $\lambda$ and $\gamma$ analysis for proposed CLRA1:}
We set parameters $\lambda=10^{\lambda^\ast}$ and $\gamma=10^{\gamma^\ast}$ for proposed CLRA1 in Eq. (\ref{eq23}) and vary both $\lambda^\ast$ and $\gamma^\ast$ from 1 to 15. Fig. \ref{figureCLRA1} shows the effect of $\lambda^\ast$ and  $\gamma^\ast$ on proposed CLRA1 with different number of microphones and sources. As can be seen from Fig.  \ref{figureCLRA1} (left and middle sub-figures), the effects of parameters $\lambda^\ast$ and  $\gamma^\ast$ on the recovery rate of CLRA1 is similar, when both $\lambda^\ast$ and  $\gamma^\ast$ are small (less than 5), the recovery rate is almost 0\%, and when  $\lambda^\ast>5$ and $\gamma^\ast\leq\lambda^\ast$, the recovery rate is larger than 0\%. This indicates that  $\lambda^\ast$ and  $\gamma^\ast$ have a joint effect on proposed CLRA1. In addition, we can see that when $\lambda^\ast=\gamma^\ast$, the peaks regarding recovery rate are achieved.

Besides, from right subfigure in Fig. \ref{figureCLRA1}, we can see that when $\lambda^\ast$ is less than 8, the recovery rate for proposed CLRA1 is almost 0\%, and when $\lambda^\ast>8$ and $\gamma^\ast \leq \lambda^\ast+2$, the recovery rate is larger than 0\%. In addition, the peaks regarding recovery rate are obtained when  $\lambda^\ast=\gamma^\ast+3$. Thus, if there is no special mention, we set the following parameters for proposed CLRA1, i.e., $\lambda^\ast=10$ and $\gamma^\ast=10$ in both C1 and C3, $\lambda^\ast=12$ and $\gamma^\ast=9$ in C2.

% \begin{figure}[!ht]
%   \centering
%   \subfigure[\(M=8\);  \(N=15\)]{\includegraphics[width=0.7\linewidth]{p13pa815.png}}
%       \subfigure[\(M=15\);  \(N=8\)]{\includegraphics[width=0.7\linewidth]{p12pa158.png}}
%     \caption{ The effect of parameters: (a) for $\beta^\ast$ on proposed CLRA2 method and (b) for $\alpha^\ast$ on proposed CLRA3 method.}
%     \label{figureCLRA2}
% \end{figure}

\textit{2) Parameter $\beta$ analysis for proposed CLRA2:} We set parameter $\lambda=10^{10}$ and analyze the effect of parameter $\beta$ on CLRA2. Denote $\beta=10^{\beta^{\ast}}$ and  we vary $\beta^{\ast}$ from 1 to 15, then Fig. \ref{fig:analysislayers} plots the recovery rate of proposed CLRA2 when $\beta^\ast$ varies. It can be seen that $\beta^\ast$ has a big effect on the performance of proposed CLRA2. In details, when $\beta^\ast$ is small (less than 8), the recovery rate is stable (about 8\%) and when $\beta^\ast$ reaches to 11, the proposed CLRA2 can achieve about 28\% recovery rate, then the recovery rate decreases once $\beta^\ast$ increases further. Thus, we set $\beta^\ast=11$ for the proposed CLRA2 if there is no special mention.

%\begin{figure}[ht]
 %  \centering
  %  \includegraphics[width=0.4\textwidth ]{p12pa158.png}
  %  \caption{The effect of parameter $\alpha^\ast$ on proposed CLRA3 method with  \(M=15\)  and  \(N=8\). }
 %   \label{figureCLRA3}
%\end{figure}

\textit{3) Parameter $\alpha$ analysis for proposed CLRA3:}  We set parameter $\lambda=10^{10}$
 and  analyze the effect of parameter $\alpha$ on CLRA3. Denote $\alpha=10^{\alpha^\ast}$ and we vary $\alpha^\ast$ from 1 to 15. It can be seen from Fig. \ref{fig:analysislayers} that the parameter $\alpha^\ast$ also has a big effect on proposed CLRA3. In details, when $\alpha^\ast<9$, the recovery rate of proposed CLRA3 is stable (about 17\%), then the recovery rate increases once the value of $\alpha^\ast$ continues to increase, and we can see that the proposed CLRA3 can reach about 37\% recovery rate when $\alpha^\ast=11$, thus, we choose $\alpha^\ast=11$ for proposed CLRA3 if no special mention.

\subsection{Comparison of the Performance for low-rank Properties}

\begin{figure*}[h]
\centering
\subfigure[$M=10$; $N=10$]{\includegraphics[trim=0.2cm 0.4cm 1.3cm 0.5cm, clip=true, width=0.3\linewidth]{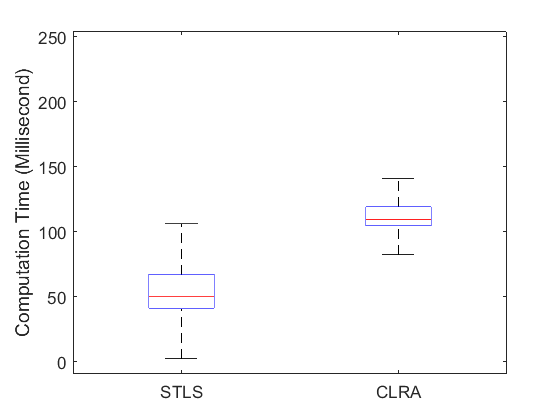}}
\subfigure[$M=15$; $N=8$]{\includegraphics[trim=0.2cm 0.4cm 1.3cm 0.5cm, clip=false,width=0.3\linewidth]{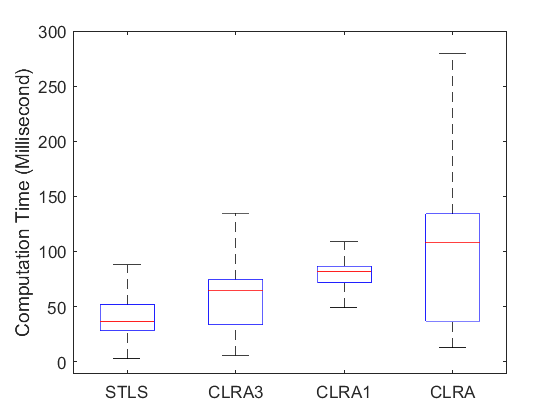}}
\subfigure[$M=8$; $N=15$]{\includegraphics[trim=0.2cm 0.4cm 1.3cm 0.5cm, clip=false,width=0.3\linewidth]{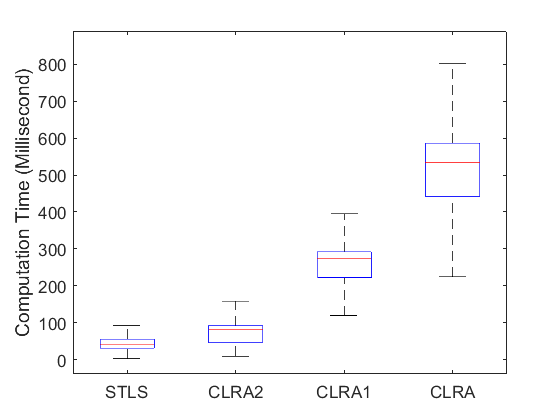}}
\caption{The running time for proposed CLRA methods in comparison with STLS.}
\label{figureTC}
\end{figure*}

This subsection shows the performance of the proposed CLRA methods in comparison with the STLS~\cite{8,39}.

The parameter $\lambda=10^{10}$ for STLS in Eq. (\ref{eq23}) (\(\gamma=\alpha=\beta=0\)). For the proposed CLAR method that combines one proposed  low-rank property with LRP, i.e.,  CLRA1, CLRA2 and CLRA3, the parameters are illustrated already in Section VII-B. For the proposed CLAR method that utilizes all the four low-rank properties, we categorize them as three cases (see Fig. \ref{figure:precodition}), 1) C3: the parameters $\lambda$ and $\gamma$ of proposed CLRA in Eq. (\ref{eq23}) ($\alpha=\beta=0$) are set to be \(10^{10}\) and \(10^{10}\), respectively.  2) C1: we set the parameters $\lambda=10^{10}$, $\gamma=10^{10}$ and $\alpha=10^{11}$ for proposed CLRA in Eq. (\ref{eq23}) ($\beta=0$). 3) C2: we set the parameters $\lambda=10^{12}$, $\gamma=10^{9}$ and $\beta=10^{13}$ for proposed CLRA in Eq. (\ref{eq23}) ($\alpha=0$). Besides, we vary both $M$ and $N$ from 1 to 15, and for each pair of fixed $M$ and $N$, we the number of configuration \(N_c(M,N)=200\) and the number of initializations $I_n(M,N)=100$ for each configuration.  

\subsubsection{Comparison of Recovery Rate}
Fig. \ref{figurecompari} shows the recovery rate of proposed CLRA methods in comparison with the STLS. First, we analyze the results of the recovery rate for STLS, 
%we can see that the STLS doesn't work when $M\leq 5$ and $N< 5$; When $M=6$, if $N\leq 14$, the recovery rate is 0\%, and if $N> 15$, the recovery rate is quite small (near 0\%, this can be seen from the \ref{figurecompari} (b)); When $N=5$ and $M<14$
it is obvious that the recovery rate achieved by STLS is about 0\% when $M\leq 6$ or $N\leq 6$; And when both $M\geq 7$ and $N\geq 7$, the recovery rate achieved by the STLS ranges from  0\% to 28\%.

%In addition, we can see a interesting phenomenon, $M$ and $N$ are symmetric, but the performance they achieved are different; for example, we can see that if $(M=20; N=12)$ and $(M=12; N=20)$,  the recovery rate achieved by  $(M=20; N=12)$ is better than the recovery rate by $(M=12; N=20)$, this can be explained by the following analysis, as can be seen from (\ref{eq23}), when $\gamma=\alpha=\beta=0$, the size of unknown matrix $X$ is $3$ by $N-1-3$, thus, if  $(M=20; N=12)$, there are $3(12-1-3)$ unknown parameters in $X$ need to be estimated; if  $(M=12; N=20)$, there are $3(20-1-3)$ unknown parameters in $X$ need to be estimated; it means that  there are $83$ more parameters need to be estimated for $(M=12; N=20)$ in comparison with the situation that $(M=20; N=12)$, therefore, we can see that the recovery rate is different for those two situations, though $M$ and $N$ are symmetric. 
Second, we compare the recovery rate of proposed CLRA1 with STLS, we can see that when both $M\leq6$ or $N\leq5$, the recovery rate obtained by the proposed CLRA1 is about 0\% which is the same as the performance of STLS; When $M=7$ and $N\geq 10$, the recovery rate achieved by proposed CLRA1  ranges from 0\% to 8\% which is better than recovery rate achived by STLS. We can also see that when $M\geq 14$ and $N= 6$, the proposed CLRA1 can obtain about 1\% recovery rate, however, the recovery rate achieved by the STLS is 0\%. In addition, when both $M>12$ and $N> 8$, the recovery rate achieved by proposed CLRA1 is about 10\% to 20\% percentage points higher than STLS. In conclusion, when  $M\geq 7$ and $N\geq 6$, the recovery rate achieved by proposed the CLRA1 is better than STLS. This validates the proposed LRPV3 and CLRA1.

\par
Third, we compare proposed CLRA2 with STLS.
%It should be noted that LRPV2 only works under the precondition $N>M+3$ and $M>4$.
When $M= 6$ and $N>13$, the proposed CLRA2 can achieve around 2\% to 4\% recovery rate which is better than the 0\% recovery rate achieved by STLS. When $M\geq 7$ and $N>M+3$, the recovery rate achieved by the proposed CLRA2 is much better than the STLS, especially when $N=15$ and $7\leq M\leq 10$, the recovery rate achieved by the proposed CLRA2 is 18\% to 27\% percentage points higher than STLS. In conclusion, when  $M\geq 6$ and $N>M+3$, the recovery rate achieved by proposed CLRA2 is better than STLS. This validates the proposed LRPV2 and CLRA2.  

\par
Fourth, we compare the proposed CLRA3 with STLS.
%It should also be noted that LRPV1 only works under the precondition $M>N+3$ and $N>4$.
 When $N= 6$ and $M>13$, the proposed CLRA3 can achieve around 2\% to 4\% recovery rate which is better than the 0\% recovery rate achieved by STLS. When $N\geq 7$ and $M>N+3$, the recovery rate achieved by proposed CLRA3 is also much better than the STLS, especially when $M=15$ and $7\leq N\leq 10$, the recovery rate achieved by proposed CLRA2 is about 12\% to 19\% percentage points higher than STLS. In conclusion, when  $N\geq 6$ and $M>N+3$, the recovery rate achieved by the proposed CLRA3 is better than STLS. This validates the proposed LRPV1 and CLRA3.  
\par
Finally, we compare the proposed CLRA method (LRP plus LRPV1, LRPV2 and LRPV3) with STLS, we can also see that the proposed CLRA method achieves a much better recovery rate than the STLS when $M>5$ and $N>5$. In addition, by combining LRP with the proposed three variants of LRP, the recovery rate achieved by the CLRA method is better than the way that just combines one proposed low-rank property (LRPV1/LRPV2/LRPV3) with LRP when $M>5$ and $N>5$.  This validates the proposed LRPV1, LRPV2 and LRPV3 again.

\subsubsection{Comparison of Convergency Rate}

Fig. \ref{figurecompari1}(a) shows the convergency rate and the corresponding percentage points between STLS and proposed CLRA methods are shown in Fig. \ref{figurecompari1}(b) for better illustration. We analyze the results of the convergency rate for STLS in Fig. \ref{figurecompari1}(a) first, 
%we can see that the STLS doesn't work when $M\leq 5$ and $N< 5$; When $M=6$, if $N\leq 14$, the recovery rate is 0\%, and if $N> 15$, the recovery rate is quite small (near 0\%, this can be seen from the \ref{figurecompari} (b)); When $N=5$ and $M<14$
we can see that the convergency rate achieved by the STLS is about 0\% when $M< 7$ or $N< 6$; And  the convergency rate achieved by STLS becomes higher when both $M$ and $N$ increase. In addition, when $M \geq 13$ and $N\geq 12$, the convergency rate achieved by the STLS is larger than 90\%.

%In addition, we can see a interesting phenomenon, $M$ and $N$ are symmetric, but the performance they achieved are different; for example, we can see that if $(M=20; N=12)$ and $(M=12; N=20)$,  the recovery rate achieved by  $(M=20; N=12)$ is better than the recovery rate by $(M=12; N=20)$, this can be explained by the following analysis, as can be seen from (\ref{eq23}), when $\gamma=\alpha=\beta=0$, the size of unknown matrix $X$ is $3$ by $N-1-3$, thus, if  $(M=20; N=12)$, there are $3(12-1-3)$ unknown parameters in $X$ need to be estimated; if  $(M=12; N=20)$, there are $3(20-1-3)$ unknown parameters in $X$ need to be estimated; it means that  there are $83$ more parameters need to be estimated for $(M=12; N=20)$ in comparison with the situation that $(M=20; N=12)$, therefore, we can see that the recovery rate is different for those two situations, though $M$ and $N$ are symmetric. 

Fig. \ref{figurecompari1}(b) shows the corresponding percentage points of convergency rate between CLRA methods and STLS.  Now, we analyze the percentage point between CLRA1 and STLS first, we can see that when  $M<6$ or $N<6$, the percentage points between CLRA1 and STLS are about 0\% since the convergency rate for both of them is 0\%. When both  $M>10$ and $N>9$, the convergency rate achieved by proposed CLRA1 is about 2\% to 18\% percentage points higher than STLS. When  $6\leq M \leq 9$ or  $5\leq N \leq 8$,  the convergency rate achieved by proposed CLRA1 is about 6\% to 58\% percentage points higher than STLS. This validates the proposed LRPV3 and CLRA1.

\par
Then, we compare the proposed CLRA2 with STLS. 
%It should also be noted that the LRPV2 only works under the precondition $N>M+3$ and $M>4$.
Since the convergency rate for both STLS and CLRA2 is 0\% when  $M= 5$, we can see that the percentage point between CLRA2 and STLS is about 0\%. When $M= 6$ and $N>12$, the convergency rate achieved by proposed CLRA2 is about 28\% to 63\% percentage points higher than STLS. When $M\geq 7$ and $N>M+3$, the convergency rate achieved by the proposed CLRA2 is also better than the STLS, especially when $M=7$ and $N>10$, the convergency rate achieved by proposed CLRA2 is about 11\% to 50\% percentage points higher than STLS. In conclusion, when  $M\geq 6$ and $N>M+3$, the convergency rate achieved by the proposed CLRA2 is better than STLS. This validates the proposed LRPV2 and CLRA2.  

\par
Next, we compare the proposed CLRA3 with STLS. 
%It should also be noted that LRPV1 only works under the precondition $M>N+3$ and $N>4$.
When $N= 6$ and $M>12$, the convergency rate achieved by proposed CLRA3 is about 32\% to 58\% percentage points higher than STLS. When $N\geq 7$ and $M>N+3$, the convergency rate achieved by proposed CLRA3 is also better than the STLS, especially 
when $N= 7$ and $M \geq 11$, the convergency rate achieved by proposed CLRA3 is about 32\% to 58\% percentage points higher than STLS. And when $N\geq 8$ and $M>N+3$, the convergency rate achieved by proposed CLRA3 is about 14\% to 19\% percentage points higher than STLS. In conclusion, when  $N\geq 6$ and $M>N+3$, the convergency rate achieved by the proposed CLRA3 is better than STLS. This validates the proposed LRPV1 and CLRA3. 
\par
Finally, we also compare the proposed CLRA method (LRP plus LRPV1, LRPV2 and LRPV3) with STLS, we can also see that proposed CLRA method achieves much better convergency rate than the STLS when $M>5$ and $N>5$. In addition, by combining all of the proposed three rank properties with LRP, the convergency rate achieved by CLRA method is also better than the way that just combine one proposed low-rank property (LRPV1/LRPV2/LRPV3) with the LRP when $M>5$ and $N>5$.  This validates the proposed LRPV1, LRPV2 and LRPV3 again.

\subsection{Computational Complexity Analysis}

\begin{figure}[t]
\centering
    \includegraphics[trim=1.8cm 0.1cm 1.8cm 0.6cm, clip=true,   width=0.4\textwidth]{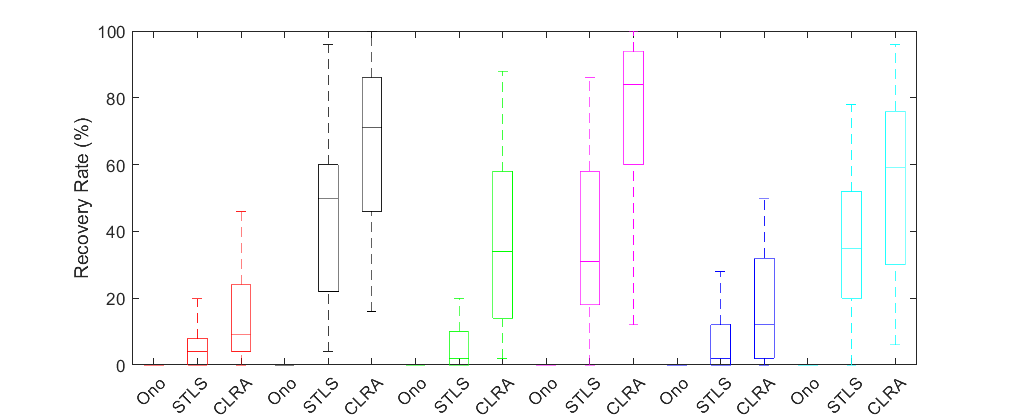}
\caption{The performance comparison in terms of recovery rate of each configuration with  C3 ($M=N=10$ (\protect\raisebox{1pt}{\protect\tikz \protect\draw[red,fill=red] (0,0) rectangle (1.ex,1.ex);}) and $M=N=18$ (\protect\raisebox{1pt}{\protect\tikz \protect\draw[
black,fill=black] (0,0) rectangle (1.ex,1.ex);})), C1 ($M=15$, $N=8$ (\protect\raisebox{1pt}{\protect\tikz \protect\draw[green,fill=green] (0,0) rectangle (1.ex,1.ex);}) and  $M=20$, $N=14$ (\protect\raisebox{1pt}{\protect\tikz \protect\draw[magenta,fill=magenta] (0,0) rectangle (1.ex,1.ex);})) and C2 ($M=8$, $N=15$ (\protect\raisebox{1pt}{\protect\tikz \protect\draw[blue,fill=blue] (0,0) rectangle (1.ex,1.ex);}) and $M=14$, $N=20$ (\protect\raisebox{1pt}{\protect\tikz \protect\draw[cyan,fill=cyan] (0,0) rectangle (1.ex,1.ex);})).}
\label{figurecomono}
\end{figure}

%\begin{figure*}[ht]
%\centering
%\subfigure[$M=10$; $N=10$]{\includegraphics[width=0.29\linewidth]{RN_1010.png}}
%\subfigure[$M=15$; $N=8$]{\includegraphics[width=0.29\linewidth]{RN_158.png}}
%\subfigure[$M=8$; $N=15$]{\includegraphics[width=0.29\linewidth]{RN_815.png}}
%\caption{The performance comparison in terms of recovery number of each configuration.}
%\label{figurecomono}
%\end{figure*}
%width=0.25\textwidth 

In this subsection, the computational complexity for the proposed CLRA method is analyzed in comparison with the STLS~\cite{8,39} that just utilizes LRP only.

Since both STLS and proposed CLAR method are based on the Gauss-Newton method, the most computation intensive part is to update the variable $p$ (see Eq. (\ref{eq31}) and Fig. \ref{Algorithm}). There are three main mathematical calculation operations in Eq. (\ref{eq31}). 1) calculate the multiplication for two Jacobian matrices, i.e., $J^{(m)^{T}} J^{(m)}$ (see \textit{Section A4 of Supplementary Material} for the form of Jacobian matrix); 2) calculate the inverse of matrix $J^{(m)^T}J^{(m)}$, i.e., $(J^{(m)^T}J^{(m)})^{-1}$; 3) calculate the multiplication for $(J^{(m)^{T}}J^{(m)})^{-1}$ and $J^{(m)^T}$. 

If only LRP is utilized for timing information estimation, there are just three sub-variables in variable $p$ that need to be estimated, i.e., $\delta$, $\eta$ and $X$, thus, the size of the corresponding Jacobian matrix is $(M-1)(2(N-1)-3)$ by $M-3^2+(3+1)(N-1)$. Now denote $(M-1)(2(N-1)-3)$ and $M-3^2+(3+1)(N-1)$ as $\hat{J}_{1,r}$ and $\hat{J}_{1,c}$, respectively. Then, we can have the computational complexity: 1) the computational complexity to calculate the multiplication for two Jacobian matrices is $O({\hat{J}}_{{1,c}^2}{\hat{J}}_{1,r})$; 2) the computational complexity to calculate the inverse of matrix $J^{(m)^T}J^{(m)}$ is $O(\hat{J}_{1,c}^3)$; 3) the computational complexity to calculate the multiplication of ${(J^{(m)^T}J^{(m)})}^{-1}$ and $J^{(m)^T}$ is also $O({\hat{J}}_{1,c}^2{\hat{J}}_{1,r})$. Thus, the computational complexity by utilizing LRP only is 
\begin{equation}
\label{tceq1}
    O(min (\hat{J}_{1,c}^2\hat{J}_{1,r}, \hat{J}_{1,c}^3)).
\end{equation}

For the proposed CLRA method, the computational complexity is analyzed with three cases according to the number of microphones and sources, i.e., C1, C2 and C3. In C3, only LRP and LRPV3 are used, and there are four sub-variables in variable $p$ that need to estimated, i.e., $\delta$, $\eta$, $X$ and $Y$, then the size of the Jacobian matrix is $(M-1)(6(N-1)-3-M_N)$ by $M-3^2-M_N^2+(3+1+2M_N)(N-1)$. Now denote $(M-1)(6(N-1)-3-M_N)$  and $M-3^2-M_N^2+(3+1+2M_N)(N-1)$ as $\hat{J}_{1,4,r}$ and $\hat{J}_{1,4,c}$, respectively. Thus, the computational complexity by utilizing the LRP and LRPV3 is 
\begin{equation}
\label{tceq2}
    O(min(\hat{J}_{1,4,c}^2\hat{J}_{1,4,r}, \hat{J}_{1,4,c}^3)). 
\end{equation}

In C1, LRP, LRPV1 and LRPV3 are used for proposed CLRA method, there are five variables that need to be estimated, i.e., $\delta$, $\eta$, $X$, $Y$ and $Z$, then the size of the Jacobian matrix is $(M-1)(7(N-1)-23-M_N)$ by $M-23^2-M_N^2+(3+N+2M_N)(N-1)$.  Now denote $(M-1)(7(N-1)-23-M_N)$  and $M-23^2-M_N^2+(3+N+2M_N)(N-1)$ as $\hat{J}_{1,2,4,r}$ and $\hat{J}_{1,2,4,c}$, respectively. Thus, the computational complexity of utilizing the LRP, LRPV1 and LRPV3 is 
 \begin{equation}
 \label{tceq3}
 O(min(\hat{J}_{1,2,4,c}^2\hat{J}_{1,2,4,r}, \hat{J}_{1,2,4,c}^3)).
 \end{equation}
 
In C2, LRP, LRPV2 and LRPV3 are used for the proposed CLRA method, there are five sub-variables in variable $p$ that need to be estimated, i.e., $\delta$, $\eta$, $X$, $Y$ and $W$, then the size of Jacobian matrix is $(M-1)(7(N-1)-3-M_N)-(N-1)3$ by $M-3^2-M_N^2+(3+1+2M_N)(N-1)+(M-1-3)(M-1+3)$.  Now denote $(M-1)(7(N-1)-3-M_N)-(N-1)3$  and $M-3^2-M_N^2+(3+1+2M_N)(N-1)+(M-1-3)(M-1+3)$ as $\hat{J}_{1,3,4,r}$ and $\hat{J}_{1,3,4,c}$, respectively. Thus, the computational complexity of utilizing the LRP, LRPV2 and LRPV3 is 
\begin{equation}
\label{tceq4}
    O(min(\hat{J}_{1,3,4,c}^2\hat{J}_{1,3,4,r}, \hat{J}_{1,3,4,c}^3)).
\end{equation}

Then the differences for those four computational complexities (from Eqs. (\ref{tceq1}) to (\ref{tceq4})) are,
\begin{equation}
\label{tceq5}
    \begin{cases}
    \hat{J}_{1,4,r}-\hat{J}_{1,r}=2(M-1)(2(N-1)-M_N) \\
    \hat{J}_{1,2,4,r}-\hat{J}_{1,4,r}=(M-1)(N-1-3) \\
     \hat{J}_{1,3,4,r}-\hat{J}_{1,4,r}=(N-1)(M-1-3) 
    \end{cases},
\end{equation}
and
\begin{equation}
\label{tceq6}
    \begin{cases}
    \hat{J}_{1,4,c}-\hat{J}_{1,c}=M_N(2(N-1)-M_N) \\
    \hat{J}_{1,2,4,c}-\hat{J}_{1,4,c}=(N-1+3)(N-1-3) \\
     \hat{J}_{1,3,4,c}-\hat{J}_{1,4,c}=(M-1+3)(M-1-3)
    \end{cases}.
\end{equation}

Upon insepction of Eqs. (\ref{tceq1}) to (\ref{tceq6}), it is obvious that with one more additional proposed rank properties (LRPV1/LRPV2/LRPV3), the computational complexity increases since the size of Jacobian matrix is larger.

Now, we show the running time of the proposed CLRA methods and also compare it with the STLS~\cite{8,39} based on three cases, i.e., C1, C2 and C3. Given $M$ microphones and $N$ sources, we set the number of configurations \(N_c(M,N)=50\), and for each configuration, we set the number of initializations  $I_n(M,N)=100$ for each configuration. Therefore, there are 5000 implementations. 
%In addition, the maximal number of iterations for each implementation is 100. 

Fig. \ref{figureTC} shows the running time of those 5000 implementations for three cases, i.e. Fig. \ref{figureTC}(a) for C3, Fig. \ref{figureTC}(b) for C1 and Fig. \ref{figureTC}(c) for C2. Fig. \ref{figureTC}(a) shows the case with $M=10$ and $N=10$, in this case, only LRP and LRPV3 can be used for proposed CLRA method. We can see with LRP only, the running time of STLS for those 5000 implementations ranges from 3 milliseconds to 110 milliseconds, which is quite fast. And with both LRP and LRPV3, the size of the Jacobian matrix of CLRA is larger than the corresponing one of STLS, thus, it needs more time to update UTIm, however, it is still very fast since the running time is less than 150 milliseconds for implementation.

Fig. \ref{figureTC}(b) shows the case with $M=15$ and $N=8$, in this case, LRP, LRPV1 and LRPV3 can be used for proposed CLRA method. We can see with LRP only, the running time of STLS is also very fast, the computational time for those 5000 implementations ranges from 3 milliseconds to 100 milliseconds. And with both LRP and LRPV1 or both LRP and LRPV3, the size of the Jacobian matrix is larger than the corresponding one of LRP only, thus, both CLRA3 and CLRA1 need more time to update the UTIm in comparison with STLS. And we can also see that the median value of CLRA1 is larger than the median value of CLRA3 due to the size of the matrix of LRPV3  is larger than the size of the matrix of LRPV1 (see Eqs. (\ref{thrm1}) and (\ref{thrm3})), resulting in more calculation time. In addition, the implementation time of CLRA is larger than both CLRA1 and CLRA3 since the size of Jacobian matrix of CLRA is bigger than the size of Jacobian matrices in both CLRA1 and CLRA3. However, it is still quite fast since the time consuming is less than 300 milliseconds for one implementation.

Fig. \ref{figureTC}(c) shows the case with $M=8$ and $N=15$,  and LRP, LRPV2 and LRPV3 can be used for proposed CLRA method in this case. We can see that the case of running time in Fig. \ref{figureTC}(c) is similar to the case in Fig. \ref{figureTC}(b), and the running time for those methods in Fig. \ref{figureTC}(c)  is less than 900 milliseconds for each implementation.
%, i.e., 1) STLS (LRP only); 2) CLRA2 (LRP and LRPV2); 3) CLRA1 (LRP and LRPV3); 4) CLRA (LRP, LRPV2 and LRPV3).

\subsection{Comparison of the Performance with Other Method}

\begin{figure}[t!]
   \centering
    \includegraphics[trim=0.7cm 0.1cm 1.0cm 0.6cm, clip=true, width=0.4\textwidth ]{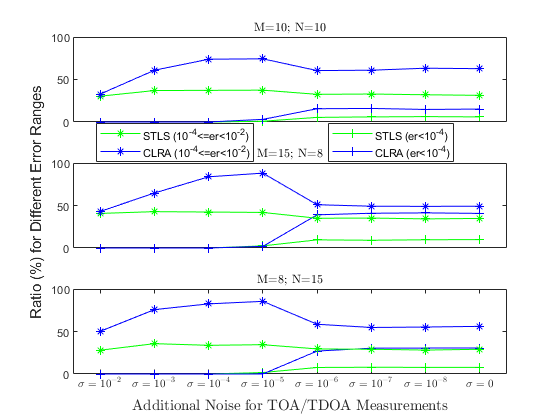}
    \caption{Two different recovery error ranges for UTIm ($er<10^{-4}s$ and $10^{-4}s\leq er <10^{-2}s$) achieved by STLS and proposed CLRA.}
    \label{simulationnoise}
    %\redcolor{add x and y axis [0, 20]}}
\end{figure}

In this subsection, we compare the proposed CLRA method with the auxiliary function-based algorithm~\cite{33}, and we name it as Ono. 

%With TOA or TDOA information, the aim for unknown timing information estimation is to obtain the distance between microphone and source, thus, after we obtain the pseudo capture time and pseudo emission time of TDOA, the distance between microphone and source can be obtained.  thus, we use the index of mean distance error to measure the performance, i.e.,
%\begin{equation}
 %   MDE=\frac{\sum_{i=1}^{M}\sum_{j=1}^{N}(\check{d}_{i,j}-d_{i,j})}{M+N}
%\end{equation}
%where $\check{d}_{i,j}$ is the estimated distance between $i^{th}$ microphone and $j^{th}$ source; and $d_{i,j}=\|r_i-s_j\|$ is the actual distance between $i^{th}$ microphone and $j^{th}$ source. In addition, we assume that the mean distance error of $10 m$ must be failed in practical applications, and thus we upper bound $MDE$ as $MDE=min(MDE, 10)$.  

The experiments are conducted with three cases, i.e., C1, C2 and C3. Given $M$ microphones and $N$ sources, we set the number of configurations \(N_c(M,N)=50\), and the number of initialization $I_n(M,N)=50$ for each configuration. Therefore, there are 2500 implementations for given $M$ and $N$.  In addition, the maximal iteration number of each implementation for Ono~\cite{33} is set to be $2 \times 10^5$. 

Fig. \ref{figurecomono} shows the recovery rate within 50 initialization for each configuration. As can be seen from Fig. \ref{figurecomono}, in all these three cases ($M=10$, $N=10$ and $M=18$, $N=18$ for C3; $M=15$, $N=8$ and $M=20$, $N=14$ for C1; $M=8$, $N=15$ and $M=14$, $N=20$ for C2), the recovery rate achieved by the Ono algorithm is always zero. And in each case, i.e., C1, C2 and C3, when both $M$ and $N$ increase, the recovery rate achieved by both STLS and CLRA continue to increase. In addition, the recovery rate achieved by STLS is better than the Ono. We can also see that for the fixed $M$ and $N$, CLRA achieves much higher recovery rate than the STLS in terms of minimal, max and median values, this verifies the proposed CLRA again.   

\subsection{Robust Analysis}
\begin{figure*}[t!]
   \centering
    \subfigure[Realistic Simulation]{\includegraphics[trim=0.7cm 0.1cm 1.0cm 0.3cm, clip=true, width=0.45\textwidth ]{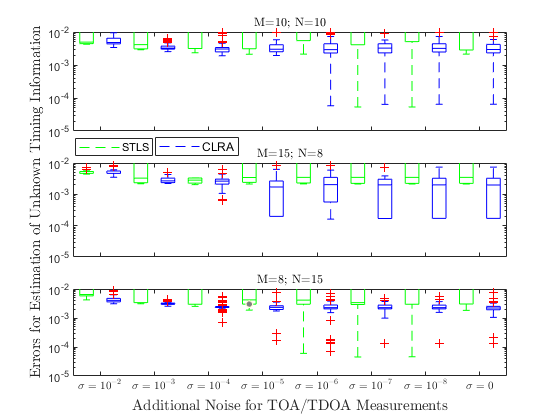}}
      \subfigure[Office Real Data] {\includegraphics[trim=0.7cm 0.1cm 1.0cm 0.3cm, clip=true, width=0.45\textwidth ]{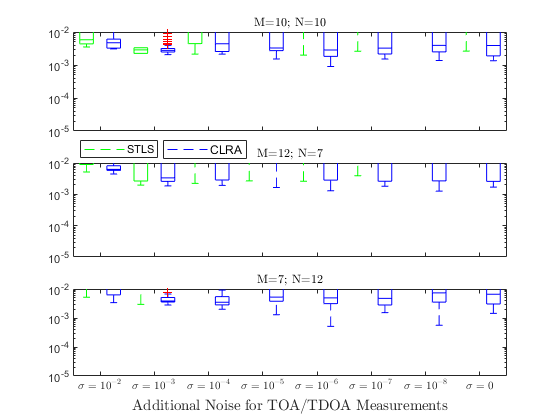}}
    \caption{Estimation errors for unknown timing information  achieved by STLS and proposed CLRA with different additional noises in TOA/TDOA measurements.}
    \label{audiosimulationnoise}
    %\redcolor{add x and y axis [0, 20]}}
\end{figure*}

In this subsection, we corrupt TOA/TDOA measurements by adding the Gaussian noise to TOA/TDOA measurements with zero mean and a standard deviation $\sigma=\{10^{-2}, 10^{-3}, \cdots, 10^{-8}\}$, thus the robustness of proposed CLRA can be shown.

First, we show the experimental results by using the TOA/TDOA measurements according to setup in Section VII-A. %We set $M=10$ and $N=10$ for C3,  $M=15$ and $N=8$ for C1, $M=8$ and $N=15$ for C2.
Both the number of configurations $N_c(M, N)=50$ and the number of initialization $I_n(M, N)=50$ are provided, it implies there are 2500 implementations for a fixed $M$ and $N$. We show  the ratio of two different estimation error ranges achieved by STLS and proposed CLRA, i.e.,  $er<10^{-4}s$ and $10^{-4}s\leq er<10^{-2}s$. When $er=10^{-4}s$, this could apply $0.034\ m$ error to the distance between mics and sources which is accurate for the task of localization. However, when $er=10^{-2}s$, this could introduce about $3.4 \ m$ distance errors that show a big impact on the task of localization. Thus, the two different error ranges could clearly depict the impact of performance achieved by STLS and CLRA on the task of localization with different noise intensity $\sigma$.  As can be seen from Fig. \ref{simulationnoise}, when the noise intensity $\sigma>10^{-6}$, the ratio for $er<10^{-4}$ achieved by both STLS and CLRA is 0. However, when noise intensity $\sigma \leq 10^{-6}$, the ratio for $er<10^{-4}$ achieved by CLRA is much higher than STLS, for example, when $M=15$ and $N=8$, the ratio $er<10^{-4}$ achieved by CLRA is about 40\% while the number achieved by STLS is just about 6\%. Besides, we can see that the ratio for $10^{-4} \leq er<10^{-2}$  achieved by CLRA is always higher than STLS, it implies that the ratio for $er>10^{-2}$ achieved by STLS is much higher than the corresponding one of CLRA. This verifies the proposed CLRA.

In addition to the TOA/TDOA simulation data above, two other different types of data are also used to conduct an robust analysis for both STLS and CLRA.  1)  Realistic Simulation: the location of mics and sources are randomly generated inside the room with size of $5m \times 5m \times 3m$, and the length of chirp source signal is $1s$, then the corresponding simulation audio signals~\cite{audiosim} are generated with $48k$ Hz mics sampling rate  and $340 m/s$ sound speed, and TOA/TDOA measurements are obtained by applying GCC-PHAT~\cite{gccp} method to those audio signals.  Due to the sampling rate of mics, the mean TOA/TDOA measurement errors are about $5 \times 10^{-6}s$ (see analysis from text between Eq. (\ref{eq33}) and Eq. (\ref{er})). 2) Real Data: the real data~\cite{real} was collected in an office with size of $5m \times 5m \times 3m$. There are 12 microphones with $96k$ Hz sampling rate are fixed  and a chirp was played by a loudspeaker from several different positions, and the available $12\times 23$ TOA/TDOA  matrix in this office real data is shown in references\footnote{This real data is available at \url{https://github.com/swing-research/xtdoa/tree/master/matlab}.}~\cite{34, real}.  Due to both sampling rate of mics and environment noises, the mean TOA/TDOA measurement errors are about $1 \times 10^{-4}s$. It implies that it is more challenge to estimate  in real application in comparison with simulation data.
We set the number of initialization $I_n(M, N)$  to be 100 for both of those two types of data.  Fig. \ref{audiosimulationnoise} shows the estimation errors $er$ for  both of those two types of data. As can be seen from Figs. \ref{audiosimulationnoise}(a) and (b), with different noise intensity $\sigma$, almost all of the $er$ achieved by both STLS and CLRA is larger than $10^{-4}s$, i.e., the distance errors between mics and sources are larger than $0.034m$, this is because the noises in TOA/TDOA measurements obtained from both realistic simulation and real data are larger than $10^{-6}s$. However, in general, the proposed CLRA achieves less estimation errors than STLS with different additional noises in both realistic simulation and real data. In summary, with the experimental results in both realistic simulation and real data, the estimation errors of  confirm the more potential of proposed CLRA in real-word applications in comparison with state-of-the-art.
\vspace{-1em}
\subsection{Limitations}
Proposed CLRA method contains three more variants of LRP in comparison with STLS method that utilize LRP only, with the three more linear constraints  formulated by our proposed  three variants of LRP, the proposed CLRA method could seek more global optimal solutions  with different initialization in comparison with STLS in both simulation data and real data. However, there are also some limitations for the proposed CLRA method: 1) proposed LRPV1 is limited with the number of microphones and sources, i.e., LRPV1 only works when $M-1>N-1+3$ and $N-1>3$. 2) proposed LRPV2 is also limited with the number of microphones and sources, i.e., LRPV2 only works when $N-1>M-1+3$ and $N-1>3$. 3) the proposed CLRA method is unable to denoise the TOA/TDOA measurements when the environment noises are introduced in TOA/TDOA measurements.

\section{Conclusion}

%{\color{red} THIS DISCUSSION IS TOO DETAILED FOR A CONCLUSION SECTION - IT SHOULD BE PLACED EARLIER IN THE PAPER} \bluecolor{I have moved them to Section IV-E and rewritten this part}

In this paper, the main focus is to estimate the  UTIm of TOA/TDOA measurements. By studying basic LRP between TOA/TDOA and UTIm, three new variants of LRP were proposed to exploit more low-rank structure rank information between UTIm and TOA/TDOA, and a proof was given for proposed three variants of LRP. Then, by utilizing the low-rank structure information revealed by proposed three variants of LRP  together with LPR for allowing to constrain the UTIm, we proposed combined low-rank approximate method to estimate UTIm. Experimental results showed better performance of proposed combined low-rank approximate method than state-of-the-arts in terms of both the recovery and convergency rate as well as estimation errors, this verified low-rank structure information exploited by proposed three variants of LRP for estimation of UTIm. 

%In addition, both the proposed three variants of LRP and CLRA also can be applied for wireless communication with radio signals if the microphones and sources are replaced with receivers and transmitters, respectively. Besides, if we regard TOA/TDOA measurements as images, unknown timing information as noises that we want to remove or features that we want to extract from images, the proposed three variants of LRP may also have potentials for other applications, such as image denoising and feature extraction.

As future work, we will try to decompose the Jacobian matrix of the proposed CLRA method as several small matrices, this could reduce the computational complexity of the proposed CLRA method. In addition, we will also investigate some other rank properties/restrictions for improving the accuracy of joint microphones and sources localization.  Besides, it is interesting to apply proposed three variants of LRP to wireless signal for synchronization of sensors and sources. 

\section*{Supplementary Materia}
\IEEEPARstart{T}{his} supplementary material provides the proof for proposed three  variants of low-rank property (LRP), and the form of the Jacobian matrix for proposed combined low-rank approximation (CLRA) method.  

The proof for three variants of LRP is based on two parts:

1) the LRP presented by state-of-the-arts~\cite{8,9,39} ((11) in the main manuscript: $rank(D+U)\leq 3$).

2) the theory of linear algebra~\cite{supp1, supp2, supp3}. By defining three matrices $E\in\mathbb{R}^{m \times n}$, $\Theta \in\mathbb{R}^{n \times h}$ and $O \in\mathbb{R}^{m \times h}$, and two column vectors $\theta\in\mathbb{R}^{n}$ and $o\in\mathbb{R}^{m}$, then we can have the corresponding linear algebra theorem~\cite{supp1, supp2, supp3}:

    \textit{Theorem 1:} Given one linear system $E\theta=o$, with coefficient matrix $E$, augmented matrix $\begin{bmatrix}
  E & o
\end{bmatrix}\in\mathbb{R}^{m \times (n+1)}$ and unknown column vector $\theta\in\mathbb{R}^{n}$, the following two items are sufficient and necessary if $m\geq n$:

\begin{itemize}
    \item If $rank(E)=rank(\begin{bmatrix}
      E & o
    \end{bmatrix})=n$, $E\theta = o$ has a unique solution, and vice versa. 
    \item If $rank(E)=rank(\begin{bmatrix}
      E & o
    \end{bmatrix})< n$, $E\theta = o$ has multiple solutions, and vice versa.
    %  \item If $rank(E)<rank(\begin{bmatrix}
   %   E & b
  %  \end{bmatrix})$, $E\theta = b$ has no solution, %and vice versa.  
\end{itemize}

Based on the \textit{Theorem 1}, we can extend one linear system to multiple linear systems by replacing those two column vectors, $\theta$ and $o$, with two matrices, $\Theta$ and $O$, respectively, then those two items in \textit{Theorem 1} are also sufficient and necessary~\cite{supp3}. Next, Sections A1, A2, and A3 show  the proof of proposed LRPV1, LRPV2, and LRPV3, respectively. Finally, Section A4 provides the derivation of the Jacobian matrix for the proposed CLRA method.

\section*{A1: Proof for Proposed LRPV1}
\begin{comment}
\begin{figure*}[!h]
   \centering
    \includegraphics[ width=1\textwidth ]{Figapdix.png}
    \caption{The four situations. }
    \label{fourfigu}
    %\redcolor{add x and y axis [0, 20]}}
\end{figure*}
\end{comment}

This section shows the proof  for the proposed LRPV1 in subsection A first, i.e., if $M-1>N-1+3$, we shall prove
\begin{equation}
\label{aandrea1}
  rank\left(T_1^\ast\right)\leq N-1+3,
\end{equation}
where $ T_1^\ast=
 \begin{bmatrix}
 D &  U
 \end{bmatrix} \in \mathbb{R}^{(M-1) \times 2(N-1)}$; $D\in \mathbb{R}^{(M-1) \times (N-1)}$; $U\in \mathbb{R}^{(M-1) \times (N-1)}$; $M$ and $N$ are the number of microphones and sources, respectively. Then, we show the details that result in $ rank\left(T_1^\ast\right)< N-1+3$ or $ rank\left(T_1^\ast\right)= N-1+3$ in subsection B.

\begin{figure}[t]
   \centering
    \includegraphics[width=0.45\textwidth ]{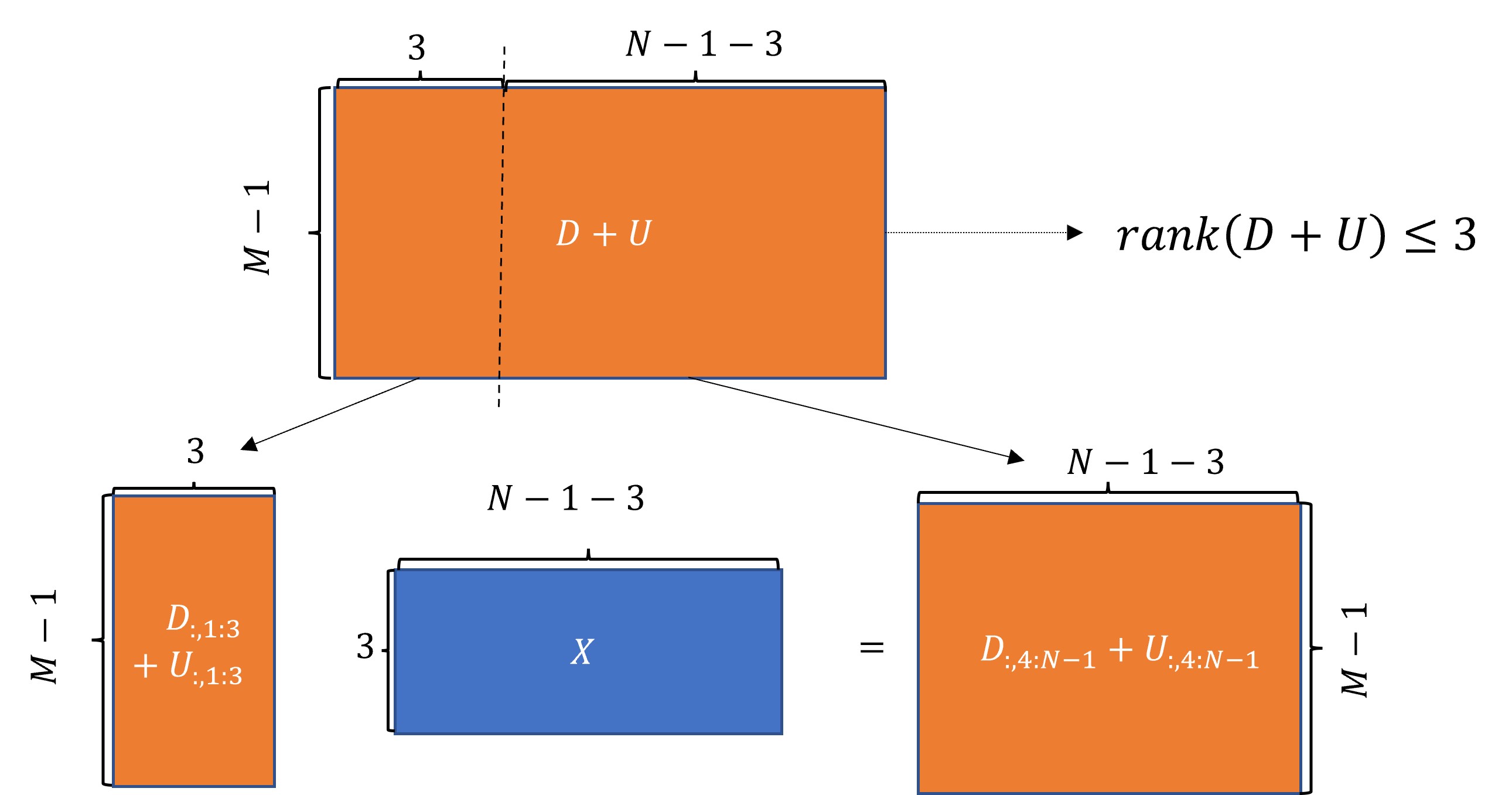}
    \caption{The illustration for the corresponding relationship of LRP with matrix $D+U$~\cite{8,9,39} ($M$: number of microphones; $N$: number of sources).}
    \label{supp1}
    %\redcolor{add x and y axis [0, 20]}}
    \vspace{-1.5em}
\end{figure}

\subsection*{A: Proof for LRPV1}
From the LRP in state-of-the-arts~\cite{8,9,39}: $rank(D+U)\leq 3$, we can see that there exist $3$ column vectors from matrix $D+U$ that could represent other column vectors of $D+U$~\cite{8}. For the convenience of analysis, we assume the first $3$ column vectors of $D+U$ are independent of each other, then we can also assume that there is an unknown matrix
\begin{equation*}
    X=
\in \mathbb{R}^{3 \times (N-1-3)},
\end{equation*}
that enables the first 3 column vectors of matrix $D+U$ to represent the remaining column vectors of matrix $D+U$ ~\cite{8,39} (see Fig. \ref{supp1}), i.e.,
\begin{equation}
    \label{anew1}
    (D_{:,1:3}+U_{:,1:3})X=D_{:,3+1:N-1}+U_{:,3+1:N-1}.
\end{equation}
 
 From Eq. (\ref{anew1}), we can have
\begin{equation}
      \label{anew2}
      D_{:,1:3}X-D_{:,3+1:N-1}+U_{:,1:3}X=U_{:,3+1:N-1}.
\end{equation}

Then, we can write Eq. (\ref{anew2}) as the matrix multiplication form
\begin{equation}
      \label{anew3}
      %
=U_{:,3+1:N-1},
\end{equation} where $I\in \mathbb{R}^{(N-1-3)\times (N-1-3)}$ is the identity matrix. 

Since matrix $%
\in \mathbb{R}^{(N-1+3)\times (N-1-3)}$, we can see that the number of row for matrix $%
$ is $N-1+3$.  In addition, we can also see that the coefficient matrix and augmented matrix in Eq. (\ref{anew3}) are $ %
$, respectively. Then based on the  \textit{Theorem 1}~\cite{supp1, supp2, supp3}, we can have
      \begin{equation}
    \label{annew11}
    rank(%
)\leq N-1+3.
\end{equation}

Now, we consider the conditions that lead matrix $%
$ to be low-rank matrix. For one matrix,   the number of both rows and columns for this matrix should be larger than the corresponding rank if such a matrix has low-rank property. Thus we need to consider two aspects, i.e., both number of columns and rows for matrix $%
 \in \mathbb{R}^{(M-1)\times 2(N-1)}$. 

1) We consider the number of columns for matrix $%
$, i.e., $2(N-1)$. Since $N-1>3$, we can have $2(N-1)>N-1+3$, which means that the number of columns for matrix $%
$ must be larger than the rank for matrix $%
$, i.e., $N-1+3$.

2) We consider the number of rows for matrix $%
$, i.e., $M-1$. From the conclusion of first aspect that the number of columns for matrix $%
$ is larger than the corresponding rank of matrix $%
$ already, we can see that if the number of rows for matrix $%
$ is larger than the rank for matrix $%
$ is a low-rank matrix. \textbf{This completes the proof for LRPV1} that  $rank\left(T_1^\ast\right)\leq N-1+3$ if $M-1>N-1+3$. 

\subsection*{B: Details for LRPV1}

\begin{figure*}[t]
   \centering
    \includegraphics[width=0.9\textwidth ]{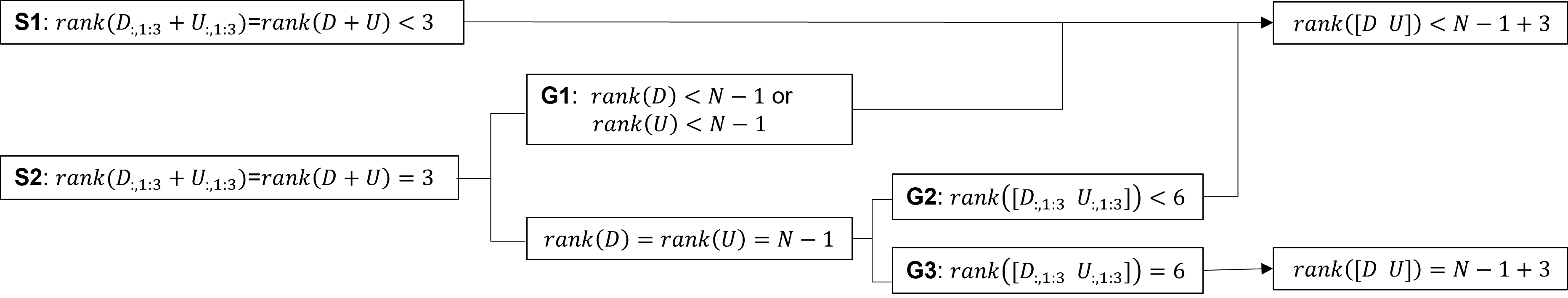}
    \caption{The steps for proving LRPV1, i.e., $rank(%
)=N-1+3$ ($N$: number of sources; S1: Situation 1; S2: Situation 2; G1: Group 1; G2: Group 2; G3: Group 3).}
    \label{SUP1}
    %\redcolor{add x and y axis [0, 20]}}
    \vspace{-1.5em}
\end{figure*}

From Eq. (\ref{annew11}), we can see that $N-1+3$ provides a upper boundary of rank for matrix $%
$ if $M-1>N-1+3$.  Now, we analyze this upper boundary, i.e., the condition that leads $rank(%
)=N-1+3$. We divide the corresponding proof into two situations, i.e., $rank(D+U)<3$ and $rank(D+U)=3$. Fig. \ref{SUP1} shows the full procedure in this subsection.

\textit{\textbf{Situation 1:}} If $rank(D_{:,1:3}+U_{:,1:3})<3$,  we can have $rank(%
)<N-1+3$.

\textit{\textbf{Proof:}} From Eq. (\ref{anew3}), we can see that sub-matrix $I$ in matrix $%
$ is identity matrix. Since the identity matrix is unique, we need to consider whether matrix $X$ is unique or not. From the LRP~\cite{8,9,39}, the corresponding relationship in Eq. (\ref{anew1}) and the \textit{Theorem 1}~\cite{supp1, supp2, supp3}, it is obvious that matrix $X$ has multiple solutions if $rank(D_{:,1:3}+U_{:,1:3})=rank(D+U)<3$, thus, with Eq. (\ref{anew3}), we can have $rank(%
)<N-1+3$.
    
\textbf{This completes the proof for \textit{\textbf{Situation 1}}.}

 \textit{\textbf{Situation 2:}} When $rank(D_{:,1:3}+U_{:,1:3})=rank(D+U)=3$,  we can have three groups:

\begin{itemize}
    \item \textit{\textbf{Group 1:}} If $rank(D)<N-1$ or $rank(U)<N-1$, it leads  
    \begin{equation}
    \label{an_supp1_1}
            rank(%
)<N-1+3. 
    \end{equation}

  \textit{\textbf{Proof:}} We use the basic knowledge of linear algebra~\cite{supp3} to prove Eq. (\ref{an_supp1_1}),  i.e., multiplying all the elements of a certain column of the matrix by one same constant and then plus them to the corresponding elements of another column,  it is obvious that the number of rank for the corresponding matrix remains unchanged~\cite{supp3}.  Thus, if we multiply all of elements of $j^{th}$ column of matrix $%
$ by 1,  then plus them to the  $\{j+N-1\}^{th}$ column of matrix $%
$, we can have
    \begin{equation}
        \label{an_supp1_2} 
        rank(%
).
        \end{equation}

Since there just three columns of matrix $D+U$ are independent of each other, i.e., $rank(D+U)=rank(D_{:,1:3}+U_{:,1:3})=3$, then we can have
   \begin{equation}
        \label{an_supp1_3} 
        rank(%
$ in Eq. (\ref{an_supp1_3}), we can see that if $rank(D)<N-1$, it means that there exists one column vector of matrix $D$ can be represented by the remaining column vectors of matrix $D$, resulting in
     \begin{equation}
        \label{an_supp1_4} 
      rank(%
) <N-1+3.
      \end{equation}
    
    With Eqs. (\ref{an_supp1_2}), (\ref{an_supp1_3}) and (\ref{an_supp1_4}), we can have
    
        \begin{equation}
        \label{an_supp1_5} 
      rank(%
) <N-1+3.
      \end{equation}

\textit{\textbf{This completes the proof that $ rank(%
) <N-1+3$ if $rank(D)<N-1$.}}

        Similarity, if we multiply all of elements of $\{j+N-1\}^{th}$ column of matrix $%
$ by 1,  then plus them to the  ${j}^{th}$ column of matrix $%
$, we can have
    \begin{equation}
        \label{an_supp1_6} 
        rank(%
).
        \end{equation}

With Eq. (\ref{an_supp1_6}), then follow the same steps as Eqs. (\ref{an_supp1_3}), (\ref{an_supp1_4}) and (\ref{an_supp1_5}), it is easily prove that if $rank(U)<N-1$, we have
\begin{equation}
        \label{an_supp1_7} 
      rank(%
) <N-1+3.
      \end{equation}
      
\textit{\textbf{This completes the proof that $ rank(%
) <N-1+3$ if $rank(U)<N-1$.}} 
    
With Eqs. (\ref{an_supp1_5}) and (\ref{an_supp1_7}), we can conclude that $ rank(%
)<N-1+3$ if $rank(D)<N-1$ or $rank(U)<N-1$.
    
\textit{\textbf{This completes the proof for Group 1.}}

 \item \textit{\textbf{Group 2:}} If $rank(D)=rank(U)=N-1$ and $rank(%
)<N-1+3. 
    \end{equation}
    
\textit{\textbf{Proof:}} First, we show one conclusion from the precondition $rank(%
)<6$ in \textit{\textbf{Group 2}}. For any index $4\leq n \leq N-1$, it is obvious that
 \begin{equation}
     \label{an1_supp1_2}
     rank(%
)+1<6+1=7$. Finally, from Eq. (\ref{an1_supp1_2}), we can conclude that
 \begin{align}
     \label{an1_supp1_21}
     rank(%
)<7.
 \end{align} 
 
Next, we use the contradiction method~\cite{supp4} to prove   Eq. (\ref{an1_supp1_1}), i.e., we assume that  $rank(%
)<N-1+3$ is wrong, then from Eq. (\ref{annew11}), we can have
       \begin{equation}
    \label{an1_supp1_3}
            rank(%
)=N-1+3. 
    \end{equation}
    
With Eq. (\ref{an1_supp1_3}), we have the following observation.

\textit{\textbf{Observation:}}For any index $4\leq n \leq N-1$, we have $ rank(%
)=7$.

\textit{\textbf{Proof:}} With Eq. (\ref{an1_supp1_3}), then for any index $4\leq n \leq N-1$, we can have
       \begin{equation}
    \label{an1_supp1_4}
            rank(%
)=4+3=7. 
    \end{equation}

By performing the elementary operation for matrix $%
$ in Eq. (\ref{an1_supp1_4}), i.e., multiply all of elements of $j^{th}$ column of matrix $%
$ by 1,  then plus them to the  $\{j+4\}^{th}$ column of matrix $%
$, then from Eq. (\ref{an1_supp1_4}), we can have
    \begin{small}
          \begin{align}
             \label{an1_supp1_5}
              &rank(%
)=7. 
          \end{align} 
    \end{small}
        
        Since $D_{:,n}+U_{:,n}$ are dependent with $D_{:,1:3}+U_{:,1:3}$, i.e., $rank(%
)=rank(D_{:,1:3}+U_{:,1:3})$, then from Eq. (\ref{an1_supp1_5}), we can have
              \begin{align}
             \label{an1_supp1_6}
             &rank(%
)=7. 
          \end{align} 
          
Also, by performing the elementary operation for matrix $%
$ in Eq. (\ref{an1_supp1_6}), i.e., multiply all of elements of $j^{th}$ column of matrix $%
$ by $-1$,  then plus them to the  $\{j+4\}^{th}$ column of matrix $%
$,  then from Eq. (\ref{an1_supp1_6}), we can have
             \begin{align}
             \label{an1_supp1_7}
             &rank(%
)=7. 
          \end{align} 

\textit{\textbf{This completes the proof for Observation.}} 

On one side, Eq. (\ref{an1_supp1_21}) validates that $rank(%
)< 7$. On the other side,  \textit{\textbf{Observation}} shows that $rank(%
)=N-1+3$, this causes a contradiction. Since Eq. (\ref{an1_supp1_21}) is correct, we can conclude $rank(%
)=N-1+3$ is wrong and  have
    \begin{equation}
        rank(%
)<N-1+3,
    \end{equation}
under the conditions that $rank(D)=rank(U)=N-1$ and $rank(%
)<6$.
 
 \textit{\textbf{This completes the proof for Group 2.}}

     \item \textit{\textbf{Group 3:}} If $rank(D)=rank(U)=N-1$ and $rank(%
)=N-1+3. 
    \end{equation}
\end{itemize}

\textit{\textbf{Proof:}} We also use the contradiction method~\cite{supp4} to prove Eq. (\ref{an2_supp1_1}) is correct, i.e., we assume $ rank(%
)=N-1+3$ in Eq. (\ref{an2_supp1_1}) is wrong, then from Eq. (\ref{annew11}), we can have
    \begin{equation}
           \label{an2_supp1_2}
         rank(%
)<N-1+3.  
    \end{equation}
    
With Eq. (\ref{an2_supp1_2}), we can have $rank(%
)<N-1+2$. Next, we will show both $rank(%
)<N-1+2$ are wrong with two steps.
    
   \textit{\textbf{Step 1:}} If $rank(%
)=N-1+2$ is correct, we have the following two observations.
    
    \textit{\textbf{Observation 1:}} The number of ranks for matrix $%
)$ from Eq. (\ref{annew11}), it is obvious that
    \begin{equation}
    \label{ob1}
         rank(%
)=N-1+2.
    \end{equation}
    
\textit{\textbf{This completes the proof for Observation 1.}}

    \textit{\textbf{Observation 2:}} The number of ranks for matrix $%
$ is constant which means it will not vary with $N-1$.

\textit{\textbf{Proof:}}
If $rank(%
)=N-1+2$,
    it implies that for any index $4\leq n\leq N-1$, we have 
\begin{equation}
       \label{an2_supp1_3}
       rank(%
)=4+2=6
\end{equation}

 Then do the elementary operation for matrix $%
$ in Eq. (\ref{an2_supp1_3}), i.e., multiply all of elements of $j^{th}$ column of matrix $%
$ by 1,  then plus them to the  $\{j+4\}^{th}$ column of matrix $%
$, we can have
\begin{align}
     \label{an2_supp1_4}
      &rank(%
)=6.
\end{align}

Since $ D_{:,n}+U_{:,n}$ can be represented by $D_{:,1:3}+U_{:,1:3}$, i.e., $rank(D_{:,1:3}+U_{:,1:3})=rank(%
)$, then from Eq. (\ref{an2_supp1_4}), we have
   \begin{align}
     \label{an2_supp1_5}
      &rank(%
)=6.
\end{align} 

Also, by performing the elementary operation for matrix $%
$ by $-1$,  then plus them to the  $\{j+4\}^{th}$ column of matrix $%
$,  then from Eq. (\ref{an2_supp1_5}), we can have
  \begin{align}
     \label{an2_supp1_6}
      &rank(%
)=6.
\end{align} 

From the precondition that $rank(%
)=6$ in \textit{\textbf{Group 3}} and Eq. (\ref{an2_supp1_6}), we can have
    \begin{equation}
        \label{an2_supp1_7}
        rank(%
)=6.
    \end{equation}
    
   Eq. (\ref{an2_supp1_7}) implies any $D_{:,n}$ can be represented by $%
$. In addition, since $n\subset \{ %
) =6.
    \end{align}

\textit{\textbf{This completes the proof for Observation 2.}} 

  On one side, \textit{\textbf{Observation 1}} shows that the number of ranks for matrix $%
$ is a function of $N-1$, i.e., the number of ranks varies with $N-1$, On another side, \textit{\textbf{Observation 2}} shows that the number of rank for matrix $%
$ is a constant, meaning the number of rank for  matrix $%
$ will not vary with $N-1$,
    this causes a contradiction.  \textit{\textbf{Thus  $rank(%
)<N-1+2$, we have the following one observation.

     \textit{\textbf{Observation 3:}} $rank(%
)<N-1+2$,  it implies for any $4\leq n \leq N-1$, we have 
    \begin{equation}
     \label{an2_supp1_81}
        rank(%
)
    <4+2=6.
    \end{equation}

By performing the elementary operation for matrix $%
$ by 1,  then plus them to the  $\{j+4\}^{th}$ column of matrix $%
$, then from Eq. (\ref{an2_supp1_81}), we can have
    \begin{small}
          \begin{align}
             \label{an2_supp1_9}
              &rank(%
)<6. 
          \end{align} 
    \end{small}

         Since $D_{:,n}+U_{:,n}$ are dependent with $D_{:,1:3}+U_{:,1:3}$, i.e., $rank(%
)=rank(D_{:,1:3}+U_{:,1:3})$, then from Eq. (\ref{an2_supp1_9}), we can have
              \begin{align}
             \label{an2_supp1_10}
             &rank(%
)<6. 
          \end{align} 
          
Also, by performing the elementary operation for matrix $%
$ by $-1$,  then plus them to the  $\{j+4\}^{th}$ column of matrix $%
$,  then from Eq. (\ref{an2_supp1_10}), we can have
             \begin{align}
             \label{an2_supp1_11}
             &rank(%
),
\end{equation}
then with  Eq. (\ref{an2_supp1_11}), it implies 
\begin{equation}
    \label{an2_supp1_13}
    rank(%
)<6.
\end{equation}
\textit{\textbf{This completes the proof for Observation 3.}} 
    
On one side, \textit{\textbf{Observation 3}} shows that the number of rank for matrix $%
)<N-1+2$, on another side, the precondition in the \textbf{\textit{Group 3}} shows that $%
)=N-1+2$ is wrong and \textit{\textbf{Step 2}} shows that $rank(%
)<N-1+2$ is wrong, thus we can conclude that $rank(%
)\leq N-1+2$ is wrong and $rank(%
)= N-1+3 $ is correct if $rank(D)=rank(U)=N-1$ and $rank(%
)=6$. \textit{\textbf{This completes the proof for Group 3.}} 

    \begin{comment}

From Eq. (\ref{new3}), we can see that sub-matrix $I$ in matrix $%
$ is identity matrix. Since the identity matrix is unique, we need to consider whether matrix $X$ is unique or not. From the LRP~\cite{3,9,39}, the corresponding relationship in Eq. (\ref{new1}) and the \textit{Theorem 1}~\cite{supp1, supp2, supp3}, we can see that matrix $X$ has multiple solutions if $rank(D_{:,1:3}+U_{:,1:3})=rank(D+U)<3$. Thus, based on the \textit{Theorem 1}~\cite{supp1, supp2, supp3} again, we can have
\begin{itemize}
    \item If $rank(D+U)=3$, matrix $X$ has a unique solution. It leads  
    \begin{equation}
    \label{supp1_1}
            rank(%
)<N-1+3. 
    \end{equation}
\end{itemize}

Besides, based on \textit{Theorem 1}~\cite{supp1, supp2, supp3}, we can also see that whatever $rank(%
)=N-1+3$, the linear systems $%
=U_{:,3+1:N-1}$ in Eq. (\ref{new3}) still hold, and this is key for unknown timing information estimation since the unknown timing information is contained in matrix $U$. 

\end{comment}

%\textbf{This completes the proof for LRPV1.}

\section*{A2: Proof for Proposed LRPV2}
This section shows the proof for proposed LRPV2 in subsection A first, i.e., if $N-1>M-1+3$, we shall prove
\begin{equation}
\label{aeq11}
    rank\left(T_2^\ast\right)\leq  M-1+3,
\end{equation}
where $ T_2^\ast=
 %
 \in \mathbb{R}^{(N-1) \times 2(M-1)}$. Then, we show the details that lead $ rank\left(T_2^\ast\right)< M-1+3$ or $ rank\left(T_2^\ast\right)= M-1+3$ in subsection B.

\subsection*{A: Proof for LRPV2}
From the LRP in state-of-the-arts~\cite{8,9,39}: $rank(D+U)=rank(D^T+U^T)\leq 3$, we can see that there exists $3$ row vectors from matrix $D+U$ that could represent remaining row vectors of $D+U$~\cite{8}. For the convenience of analysis, we assume the first $3$ row vectors of matrix $D+U$ are independent from each other, then we can assume that there is an unknown matrix
\begin{equation*}
    \hat{X}= \begin{bmatrix}
\hat{X}_{1,1} & \cdots & \hat{X}_{1,M-1-3} \\
\hat{X}_{2,1} & \cdots & \hat{X}_{2,M-1-3} \\
   \hat{X}_{3,1} & \cdots &  \hat{X}_{3,M-1-3} 
\end{bmatrix}\in \mathbb{R}^{3 \times (M-1-3)},
\end{equation*}
that enables the first 3 row vectors of matrix $D+U$ to represent the other row vectors of matrix $D+U$ (see Fig. \ref{supp2}) ~\cite{8,39}, i.e.,
\begin{equation}
    \label{anew1_1}
    (D_{1:3,:}^T+U_{1:3,:}^T)\hat{X}=D_{3+1:M-1,:}^T+U_{3+1:M-1,:}^T.
\end{equation}

\begin{figure}[t]
   \centering
    \includegraphics[width=0.45\textwidth ]{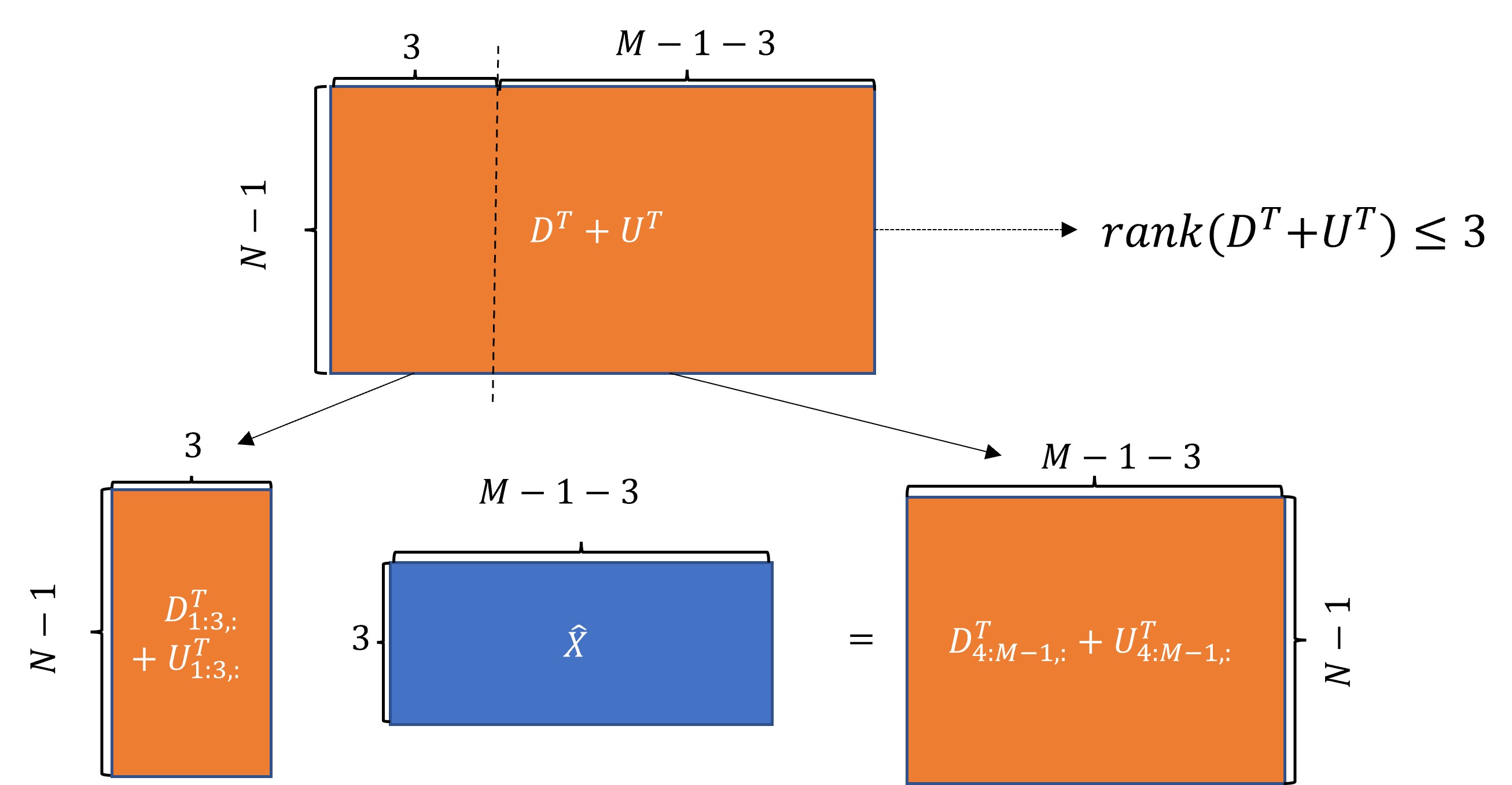}
    \caption{The illustration for the corresponding relationship of LRP with matrix $D^T+U^T$~\cite{8,9,39} ($M$: number of microphones; $N$: number of sources).}
    \label{supp2}
    %\redcolor{add x and y axis [0, 20]}}
    \vspace{-1.5em}
\end{figure}

From Eq. (\ref{anew1_1}), we can have
\begin{equation}
      \label{anew2_1}
      D_{1:3,:}^T\hat{X}-D_{3+1:M-1,:}^T+U_{1:3,:}^T\hat{X}=U_{3+1:M-1,:}^T.
\end{equation}

Then, we can write Eq. (\ref{anew2_1}) as the matrix multiplication form
\begin{equation}
      \label{anew3_1}

=U_{3+1:M-1,:}^T,
\end{equation} where $I\in \mathbb{R}^{(M-1-3)\times(M-1-3)}$ is the identity matrix.

%Denote $%
$ by $W$, equ (\ref{new3_1}) can be written as,
%\begin{equation}
%      \label{new4_1}
%      %
\in \mathbb{R}^{(M-1+3)\times (M-1-3)}$, we can see that the number of rows for matrix $%
$ is $M-1+3$. In addition, we can also see the coefficient matrix and augmented matrix in Eq. (\ref{anew3_1}) are $ %
$, respectively. Then based on the  \textit{Theorem 1}~\cite{supp1, supp2, supp3}, we can have
\begin{equation}
    \label{annew11_1}
    rank(%
)\leq M-1+3.
\end{equation}

Now, we consider the conditions that lead matrix $%
$ to be low-rank matrix.  For one matrix, the number of both rows and columns for this matrix should be larger than the corresponding rank if such a matrix has low-rank property. Thus two aspects need to be considered, i.e., both number of columns and rows for matrix $%
\in \mathbb{R}^{(N-1)\times 2(M-1)}$. 

1) We consider the number of columns for matrix $%
$, i.e., $2(M-1)$. Since $M-1>3$, we can have $2(M-1)>M-1+3$, which means that the number of columns for matrix $%
$ must be larger than the rank  for matrix $%
$, i.e., $M-1+3$.

2) We consider the number of rows for matrix $%
$, i.e., $N-1$. From the conclusion of first aspect that the number of columns for matrix $%
$ is larger than the corresponding rank of matrix $%
$ already, we can see that if the number of rows for matrix $%
$ is larger than the rank for matrix $%
$ is a low-rank matrix.  \textbf{This completes the proof for LRPV2} that  $rank\left(T_2^\ast\right)\leq M-1+3$ if $N-1>M-1+3$.

\subsection*{B: Details for LRPV2}
From Eq. (\ref{annew11_1}), we can see that $M-1+3$ provides a upper boundary of rank for matrix $%
$ if $N-1>M-1+3$. Now, we analyze this upper boundary, i.e., the condition that leads $rank(%
)=M-1+3$. We divide the corresponding proof into two situations, i.e., $rank(D^T+U^T)<3$ and $rank(D^T+U^T)=3$.

\textit{\textbf{Situation 1:}} If $rank(D_{1:3,:}^T+U_{1:3,:}^T)=rank(D^T+U^T)<3$,  we can have $rank(%
)<M-1+3$.
    
    \textit{\textbf{Situation 2:}} If $rank(D_{1:3,:}^T+U_{1:3,:}^T)=rank(D^T+U^T)=3$,  we can have three groups.

\begin{itemize}
    \item \textit{\textbf{Group 1:}} If $rank(D^T)<M-1$ or $rank(U^T)<M-1$, it leads  
    \begin{equation}
    \label{av2-n_supp1_1}
            rank(%
)=M-1+3. 
    \end{equation}
    \end{itemize}

    \textit{\textbf{Proof:}} By using the same proof procedure as LRPV1 in Section A1-B, it is easy to prove those two situations above.
    
\textit{\textbf{This completes the proof for the conditions that lead $rank(%
)=M-1+3$.}}

\section*{A3: Proof for Proposed LRPV3}
This section shows the proof for the proposed LRPV3 in subsection A first, i.e.,
\begin{equation}
\label{aeq13}
  rank\left(T_3^\ast\right)\leq min(N-1+3,M-1+3),
\end{equation}
where $T_3^\ast= %
\in \mathbb{R}^{2(M-1) \times 2(N-1)}$. Then, we show the details that lead $ rank\left(T_3^\ast\right)< min(N-1+3,M-1+3)$ or $ rank\left(T_3^\ast\right)= min(N-1+3,M-1+3)$ in subsection B.
 
 \subsection*{A: Proof for LRPV3}
 We consider two situations for LRPV3, i.e., $M\geq N$ and $M<N$.

If $M\geq N$, the proof for LRPV3 in Eq. (\ref{aeq13}) is transformed to
\begin{equation}
\label{aeq13_1}
  rank\left(%
\right)\leq N-1+3.
\end{equation}

From Eq. (\ref{anew1}), we can have
\begin{equation}
\label{aeq14_1}
\begin{cases}
        D_{:,1:3}X-D_{:,3+1:N-1}+U_{:,1:3}X=U_{:,3+1:N-1} \\
         U_{:,1:3}X-U_{:,3+1:N-1}+D_{:,1:3}X=D_{:,3+1:N-1}
\end{cases}.
\end{equation}

Then we can write Eq. (\ref{aeq14_1}) as matrix multiplication form

\begin{equation}
\label{aeq15_1}
    \begin{cases}
             %
=D_{:,3+1:N-1}
    \end{cases},
\end{equation}where $I \in \mathbb{R}^{(N-1-3) \times (N-1-3)}$ is the identity matrix.

%Denote $%
$ by $Y$ where $Y\in \mathbb{R}^{(N-1+3) \times (N-1-3)}$, then equ (\ref{eq15_1}) can be written as,
Then Eq. (\ref{aeq15_1}) can be written as
\begin{equation}
\label{aeq16_1}
     %
\in \mathbb{R}^{(N-1+3)\times (N-1-3)}$, we can see that the number of rows for matrix $%
$ is $N-1+3$. In addition, we can see the coefficient matrix and the augmented matrix in Eq. (\ref{aeq16_1}) are $ %
$, respectively. Then based on the  \textit{Theorem 1}~\cite{supp1, supp2, supp3}, we can have
\begin{equation}
\label{aeq17_1}
    rank(%
) \leq N-1+3.
\end{equation}
    \textbf{This completes the proof for LRPV3 when $M\geq N$.}

      If $M<N$, the proof for LRPV3 in Eq. (\ref{aeq13}) is transformed to
\begin{equation}
\label{aeq13_2}
  rank\left(%
\right)\leq M-1+3.
\end{equation}

From Eq. (\ref{anew1_1}), we can have
\begin{equation}
\label{aeq14_2}
\begin{cases}
         D_{1:3,:}^T\hat{X}-D_{3+1:M-1,:}^T+U_{1:3,:}^T\hat{X}=U_{3+1:M-1,:}^T \\
          U_{1:3,:}^T\hat{X}-U_{3+1:M-1,:}^T+D_{1:3,:}^T\hat{X}=D_{3+1:M-1,:}^T
\end{cases}.
\end{equation}

Then we can write Eq. (\ref{aeq14_2}) as matrix multiplication form
\begin{equation}
\label{aeq15_2}
    \begin{cases}
             %
=D_{3+1:M-1,:}^T
    \end{cases},
\end{equation}where $I\in \mathbb{R}^{(M-1-3) \times (M-1-3)}$ is the identity matrix.

%Denote $%
$ by $\hat{Y}$ where $\hat{Y}\in \mathbb{R}^{(M-1+3) \times (M-1-3)}$.
  
Then Eq. (\ref{aeq15_2}) can be written as
\begin{equation}
\label{aeq16_2}
     %
\in \mathbb{R}^{(M-1+3)\times (M-1-3)}$, we can see that the number of rows for matrix $%
$ is $M-1+3$. In addition, we can also see  the coefficient matrix and the augmented matrix in Eq. (\ref{aeq16_2}) is $ %
$, respectively. Then based on the  \textit{Theorem 1}~\cite{supp1, supp2, supp3}, we can have
\begin{equation}
\label{aeq17_2}
    rank(%
^T$, we can have
      \begin{equation}
\label{aeq18_2}
    rank(%
) \leq M-1+3.
\end{equation}
  \textbf{This completes the proof for LRPV3 when $M< N$.}

      Finally, Based on Eq. (\ref{aeq17_1}) when $M\geq N$ and Eq. (\ref{aeq18_2}) when $M< N$, we can have
  \begin{equation}
\label{aeq18_3}
  rank(%
)\leq min(N-1+3,M-1+3).
\end{equation}
   \textbf{This completes all the proof for LRPV3}.

    \subsection*{B: Details for LRPV3}

     We first show the details of LRPV3 in Eq. (\ref{aeq17_1}) when $M\geq N$ and then the details of LRPV3 in Eq. (\ref{aeq17_2}) are displayed when $M<N$.

(1) $M\geq N$: From Eq. (\ref{aeq17_1}), we can see that $N-1+3$ provides a upper boundary of rank for matrix $%
$ if $M\geq N$.  Now, we analyze this upper boundary, i.e., the condition that leads $rank(%
)=N-1+3$.  We divide the corresponding proof into two situations, i.e., $rank(D+U)<3$ and $rank(D+U)=3$.

\textit{\textbf{Situation 1:}} If $rank(D_{:,1:3}+U_{:,1:3})=rank(D+U)<3$,  we can have $rank(%
)<M-1+3$.
    
    \textit{\textbf{Situation 2:}} If $rank((D_{:,1:3}+U_{:,1:3})=rank(D+U)=3$,  we can have three groups.

\begin{itemize}
    \item \textit{\textbf{Group 1:}} If $rank(D)<N-1$ or $rank(U)<N-1$, it leads  
    \begin{equation}
    \label{av3-1_supp1_1}
            rank(%
)=N-1+3. 
    \end{equation}
    \end{itemize}

    \textit{\textbf{Proof:}} First, by performing the elementary operation for matrix $%
$, i.e., multiply all of elements of $i^{th}$ row of matrix $%
$ by 1,  then plus them to the  $\{i+M-1\}^{th}$ column of matrix $%
$, then from Eq. (\ref{aeq17_1}), we can have
          \begin{align}
          \label{aV3-new1}
              &rank(%
)\leq N-1+3. 
          \end{align}

Then  by using the same proof procedure as LRPV1 in Section A1-B for matrix $%
$ in Eq. (\ref{aV3-new1}), it is easy to prove those two situations above.
    
\textit{\textbf{This completes the proof for the conditions that lead $rank(%
)=N-1+3$ when $M\geq N$.}}

\begin{comment}
From equ (\ref{eq16_1}), we can see that sub-matrix $I$ in matrix $%
$ is identity matrix which is unique. From the analysis for LRPV1 in Section A1 that if $rank(D_{:,1:3}+U_{:,1:3})=rank(D+U)=3$, $X$ is unique, otherwise, $X$ has multiple solutions, then based on the \textit{Theorem 1}~\cite{supp1, supp2, supp3} again, we can have
\begin{itemize}
    \item If $rank(D+U)=3$, matrix $X$ has a unique solution. It leads  
    \begin{equation}
    \label{supp1_1}
         rank(%
)<N-1+3. 
    \end{equation}
\end{itemize}
\end{comment}

(2) $M<N:$  From Eq. (\ref{aeq17_2}), we can see that $M-1+3$ provides a upper boundary of rank for matrix $%
$ if $M< N$.  Now, we analyze this upper boundary, i.e., the conditions that leads $rank(%
)=M-1+3$.  We divide the corresponding proof into two situations, i.e., $rank(D^T+U^T)<3$ and $rank(D^T+U^T)=3$.

\textit{\textbf{Situation 1:}} If $rank(D_{1:3,:}^T+U_{1:3,:}^T)=rank(D^T+U^T)<3$,  we can have $rank(%
)<M-1+3$.
    
    \textit{\textbf{Situation 2:}} If $rank(D_{1:3,:}^T+U_{1:3,:}^T)=rank(D^T+U^T)=3$,  we can have three groups.

\begin{itemize}
    \item \textit{\textbf{Group 1:}} If $rank(D^T)<M-1$ or $rank(U^T)<M-1$, it leads  
    \begin{equation}
    \label{av3-1_supp1_11}
            rank(%
)=M-1+3. 
    \end{equation}
    \end{itemize}

    \textit{\textbf{Proof:}} First, by performing the elementary operation for matrix $%
$, i.e., multiply all of elements of $j^{th}$ row of matrix $%
$ by 1,  then plus them to the  $\{j+N-1\}^{th}$ column of matrix $%
$, then from Eq. (\ref{aeq17_1}), we can have
          \begin{align}
          \label{aV3-new2}
              &rank(%
)\leq M-1+3. 
          \end{align}

Then  by using the same proof procedure as LRPV1 in Section A1-B for matrix $%
$ in Eq. (\ref{aV3-new2}), it is easy to prove those two situations above when $M<N$.

    \textit{\textbf{This completes the proof for the conditions that lead $rank(%
)=M-1+3$ when $M< N$.}}

\begin{comment}

From Eq. (\ref{eq16_2}), we can see that sub-matrix $I$ in matrix $%
$ is identity matrix which is unique. In addition, from the analysis for LRPV2 in Section A2 that if $rank(D_{1:3,:}^T+U_{1:3,:}^T)=rank(D^T+U^T)=3$, $\hat{X}$ is unique, otherwise, $\hat{X}$ has multiple solutions, thus based on the \textit{Theorem 1}~\cite{supp1, supp2, supp3} again, we can have
\begin{itemize}
    \item If $rank(D^T+U^T)=3$, matrix $\hat{X}$ has a unique solution. It leads  
    \begin{equation}
    \label{supp2_1}
           rank(%
)<M-1+3. 
    \end{equation}
\end{itemize}

 Besides, based on Eqs. (\ref{supp1_1}), (\ref{supp1_2}), (\ref{supp2_1}) and (\ref{supp2_2}), we can see that proposed LRPV3 in (\ref{eq18_3}) contains two situations
\begin{itemize}
    \item If $rank(D+U)=3$, it leads  
    \begin{equation}
    \label{supp3_1}
         rank(%
)<min \{N-1+3, M-1+3\}.
    \end{equation}
\end{itemize}
\end{comment}

\section*{A4: Form of the Jacobian matrix for CLRA Method}
This section shows the form of the Jacobian matrix of the proposed CLRA method: $J=\partial q/\partial p$
where column vector $p= %
^T$; $P=M+N-1+3(N-1-3)+M_N(2(N-1)-M_N)+(N-1+3)(N-1-3)+(M-1+3)(M-1-3)$; $M_N=min(M-1+3, N-1+3)$ 
and 
\begin{equation*}
\label{alesio_12}
\begin{cases}
      x=v(X) \\
      y=v(Y) \\
      z=v(Z) \\
      w=v(W)
\end{cases},
\end{equation*}
and   $v(\cdot)$ denotes operation for column-wise matrix vectorization; $X \in \mathbb{R}^{3 \times (N-1-3)}$; $Y \in \mathbb{R}^{M_N \times (2(N-1)-M_N})$; $Z \in \mathbb{R}^{(N-1+3) \times (N-1-3)}$; $W \in \mathbb{R}^{(M-1+3) \times (M-1-3)}$ and column vector $ q=%
\in \mathbb{R}^{Q},
% \end{equation} 
and $Q=(M-1)(8(N-1)-2M_N-6)-3(N-1)$;
%\begin{align}
%\label{alessio_123}
  %   & M_N=min(M-1+3, N-1+3), \nonumber \\
   %  & D_{i-1,j-1}=\dot{t}_{i,j}^2-\dot{t}_{i,1}^2-\dot{t}_{1,j}^2+\dot{t}_{1,1}^2, \nonumber\\
  %   &   U_{i-1,j-1}=2\dot{\delta}_i\left(\dot{t}_{i,j}-\dot{t}_{i,1}\right)-2\dot{\delta}_1\left(\dot{t}_{1,j}-\dot{t}_{1,1}\right) \nonumber \\
  %     & -2\dot{\eta}_j\left(\dot{t}_{i,j}-\dot{t}_{1,j}\right)+2\dot{\eta}_j\left(\dot{\delta}_1-\dot{\delta}_i\right), 
%\end{align}
%for $i$ and $j$ range from $2$ to $M$ and from $2$ to $N$, respectively; and 
$A=D_{:,1:3}\in \mathbb{R}^{(M-1) \times 3}$; $B=D_{:,3+1:N-1}\in \mathbb{R}^{(M-1) \times (N-1-3)}$; $F=U_{:,1:3}\in \mathbb{R}^{(M-1) \times 3}$; $G=U_{:,3+1:N-1}\in \mathbb{R}^{(M-1) \times (N-1-3)}$; $T_{11}^\ast={T_{1}^\ast}_{:,1:N-1+3}\in \mathbb{R}^{(M-1) \times (N-1+3)}$; $T_{12}^\ast={T_{1}^\ast}_{:,N+3:2(N-1)}\in \mathbb{R}^{(M-1) \times (N-1-3)}$; $T_{21}^\ast={T_{2}^\ast}_{:,1:M-1+3}\in \mathbb{R}^{(N-1) \times (M-1+3)}$; $T_{22}^\ast={T_{2}^\ast}_{:,M+3:2(M-1)}\in \mathbb{R}^{(N-1) \times (M-1-3)}$;
    $T_{31}^\ast={T_{3}^\ast}_{:,1:M_N}\in \mathbb{R}^{2(M-1) \times M_N}$; $T_{32}^\ast={T_{3}^\ast}_{:,M_N+1:2(N-1)}\in \mathbb{R}^{2(M-1) \times (2(N-1)-M_N)}$ and $\lambda$, $\alpha$, $\beta$ and $\gamma$ are parameters for LRP and proposed three variants of LRP, respectively; In addition, matrices $D$ and $U$ are defined in (9) of the main manuscript;  matrices $T_1^\ast$, $T_2^\ast$ and $T_3^\ast$ are defined in (12), (13) and (14) of the main manuscript, respectively.

The Jacobian matrix $J=\frac{\partial q}{\partial p}$ can be calculated as follows:
      \begin{equation}
          J=
.
              \label{aeq113}
     \end{equation}  
     
   \begin{comment}
     \begin{equation}
      \label{eq113}
          =
          %
.
      \end{equation}
   \end{comment}
 %$ q=%
^T$,
%\begin{equation*}
 %   \label{alesio_11}
 %   \begin{cases}
 %            f_A=v(U)\\
  %           f_B=v((A +F)X-B-G) \\
  %           f_C=v(T_{31}^\ast Y-T_{32}^\ast)  \\
  %           f_D=v(T_{21}^\ast W-T_{22}^\ast) \\
 %            f_E= v(T_{11}^\ast Z-T_{12}^\ast)
  %  \end{cases}
%\end{equation*}
%and

      Then the computation of block matrices in Eq. (\ref{aeq113}) can be expressed as follows:
      \begin{equation*}
       \label{eq114}
       \begin{cases}
           \frac{\partial f_A}{\partial\dot{\delta}}=
          %
$.

\vfill

\end{document}